\documentclass[xcolor=dvipsnames]{llncs}
\usepackage[utf8]{inputenc}
\usepackage[english]{babel}

\usepackage{listingsutf8}
\usepackage{chngcntr}
\usepackage[table,dvipsnames]{xcolor}
\usepackage{makecell}
\usepackage{longtable}
\usepackage{verbatim}
\usepackage{comment}
\usepackage{multicol}
\usepackage{subcaption}
\usepackage{makecell}
\usepackage{floatrow}
\usepackage{wrapfig}
\usepackage{rotating}
\usepackage{amsmath}
\usepackage{enumitem}
\usepackage{mathptmx}
\usepackage{syntax}
\usepackage{tikz}
\usetikzlibrary{mindmap}

\usepackage{varwidth}

\newcommand*\circled[1]{\tikz[baseline=(char.base)]{
		\node[shape=circle,draw,inner sep=1pt] (char) {#1};}}
\usepackage{tkz-euclide}
\usepackage{tikz}
\usepackage{float}
\usepackage{pgfplots}
\makeatletter
\def\addlegendimage{\pgfplots@addlegendimage}
\makeatother

\usepackage{amsthm}

\usetikzlibrary{shapes}
\usetikzlibrary{shapes.geometric} % required for the ellipse shape
\usetikzlibrary{arrows, backgrounds, calc, hobby, positioning,shadows}
\usetikzlibrary{mindmap,backgrounds,trees,matrix, positioning}
\usetikzlibrary{svg.path}
\tikzset{vertex style/.style={
		draw=#1,
		thick,
		fill=#1!70,
		text=white,
		ellipse,
		font=\small,
		outer sep=3pt, % the usage of this option will be clear later on
	},
}
\usepackage{tikz}
\usepackage{mathpazo}
\newcounter{row}
\newcounter{col}

\usepackage[final,
bookmarks=true,
hypertexnames=true,
hyperindex=true,
hyperfootnotes=true,
linktocpage=false,
hyperfigures=true,
breaklinks=true, 
pdfborder={0 0 0},
colorlinks=true,
linkcolor=black,
unicode=true,						
citecolor=blue,
urlcolor=blue,
pdfinfo={
	Title={GeoSPARQL+: Syntax, Semantics and System for Integrated Querying of Graph, Raster and Vector Data},
	Author={Timo Homburg, Steffen Staab, Daniel Janke},
	Keywords={GeoSPARQL, Raster data, Vector data, POSTGIS}            
}]{hyperref}
\usepackage[noabbrev,nameinlink]{cleveref}

\definecolor{gray}{rgb}{0.4,0.4,0.4}
\definecolor{darkblue}{rgb}{0.0,0.0,0.6}
\definecolor{cyan}{rgb}{0.0,0.6,0.6}
\theoremstyle{definition}
\lstdefinelanguage{sparql}{
	basicstyle=\scriptsize,
	morekeywords={SELECT,AS,CONSTRUCT,BIND,DESCRIBE,ASK,WHERE,FROM,NAMED,PREFIX,BASE,OPTIONAL,FILTER,GRAPH,LIMIT,OFFSET,SERVICE,UNION,EXISTS,NOT,BINDINGS,MINUS,a,geo,rdf,geo2,uom,ex,geof,school,hospital,firebrigade,dangerzone,min},
	frame=single,
	sensitive=true
}
\lstdefinelanguage{XML}
{
	basicstyle=\scriptsize,
	morestring=[b]",
	morestring=[s]{>}{<},
	morecomment=[s]{<?}{?>},
	stringstyle=\color{black},
	identifierstyle=\color{darkblue},
	keywordstyle=\color{cyan},
	captionpos=b,
	frame=single,
	morekeywords={xmlns,version,type}% list your attributes here
}
\newcommand\pfun{\mathrel{\ooalign{\hfil$\mapstochar\mkern5mu$\hfil\cr$\to$\cr}}}
\DeclareMathOperator{\dom}{dom}

\DeclareMathOperator{\var}{var}
\DeclareMathOperator{\tp}{\mbox{\emph{tp}}}

\DeclareMathOperator{\GL}{\mbox{\emph{GL}}}
\DeclareMathOperator{\BOOL}{\mbox{\emph{BOOL}}}
\DeclareMathOperator{\Geo}{\mbox{\emph{Geo}}}

\DeclareMathOperator{\RL}{\mbox{\emph{RL}}}

\DeclareMathOperator{\rastervaleq}{rastervaleq}
\DeclareMathOperator{\geometrytopointset}{geom2pset}
\DeclareMathOperator{\pointsettogeometry}{pset2geom}

\DeclareMathOperator{\STR}{STR}
\DeclareMathOperator{\Rect}{Rect}

\DeclareMathOperator{\geoequals}{geo2:equals}
\DeclareMathOperator{\geometryintersection}{geometryIntersection}
\DeclareMathOperator{\coverageintersection}{rasterIntersection}
\DeclareMathOperator{\rasterintersection}{rasterIntersection}
\DeclareMathOperator{\intersects}{geo2:intersects}
\DeclareMathOperator{\buffer}{geof:buffer}
\DeclareMathOperator{\distance}{geof:distance}
\DeclareMathOperator{\geointersects}{geof:intersects}
\DeclareMathOperator{\geointersection}{geof:intersection}
\DeclareMathOperator{\geofequals}{geof:equals}
\DeclareMathOperator{\uommeter}{uom:meter}

\DeclareMathOperator{\coverageplus}{geo2:rasterPlus}

\DeclareMathOperator{\coveragesmaller}{geo2:rasterSmaller}

\DeclareMathOperator{\rastertogeom}{raster2geom}
\DeclareMathOperator{\geomtoraster}{geom2raster}

\DeclareMathOperator{\cellval}{cellval}

\usepackage{stmaryrd} % \llbracket \rrbracket
\newcommand{\eval}[2]{\llbracket #1 \rrbracket_{#2}}

\newcommand{\condeval}[2]{#2 \models #1}
\usepackage{amssymb} % required for real number R

\usepackage{stackengine,scalerel}
\def\dclesize{\ThisStyle{\scalebox{0.8}{$\SavedStyle\bigcirc$}}}
\def\dcle{\ensurestackMath{\stackon[0pt]{\scalebox{0.8}{<}}{\dclesize}}}
\def\circledless{\def\stacktype{L}\mathbin{\scalerel*{\dcle}{\dclesize}}}
\begin{document}
\counterwithout{lstlisting}{chapter}
	
\author{Timo Homburg \inst{1}, Steffen Staab\inst{3}, Daniel Janke\inst{2}}
%\author{}
%\institute{}
\institute{
	\begin{tabular}{cc}
	timo.homburg@hs-mainz.de & steffen.staab@ipvs.uni-stuttgart.de \\
		\inst{1}Mainz University Of Applied Sciences, DE &  \inst{3}Universit\"a{}t Stuttgart, DE, and \\
			danijank@uni-koblenz.de & WAIS Research Group, \\ \inst{2}Universit\"a{}t Koblenz, DE& University of Southampton, UK\\
\end{tabular}
}
		\title{GeoSPARQL+: Syntax, Semantics and System for Integrated Querying of Graph, Raster and Vector Data}
	
		\maketitle
\vspace{-0.7cm}		

		\begin{abstract}
			We introduce an approach to semantically represent and query raster data in a Semantic Web graph. We extend the GeoSPARQL vocabulary and query language to support raster data as a new type of geospatial data. We define new filter functions and illustrate our approach using several use cases on real-world data sets. Finally, we describe a prototypical implementation and validate the feasibility of our approach.			
		\end{abstract}
	\begin{comment}
		\end{comment}
	
%\vspace{-0.9cm}
		\pagenumbering{gobble}
		
		\section{Introduction}
\label{sec:introduction}
\begin{figure}[H]
	\vspace{-0.5cm}
	\centering
\begin{subfigure}{.48\textwidth}
	\centering
	\includegraphics[width=\linewidth]{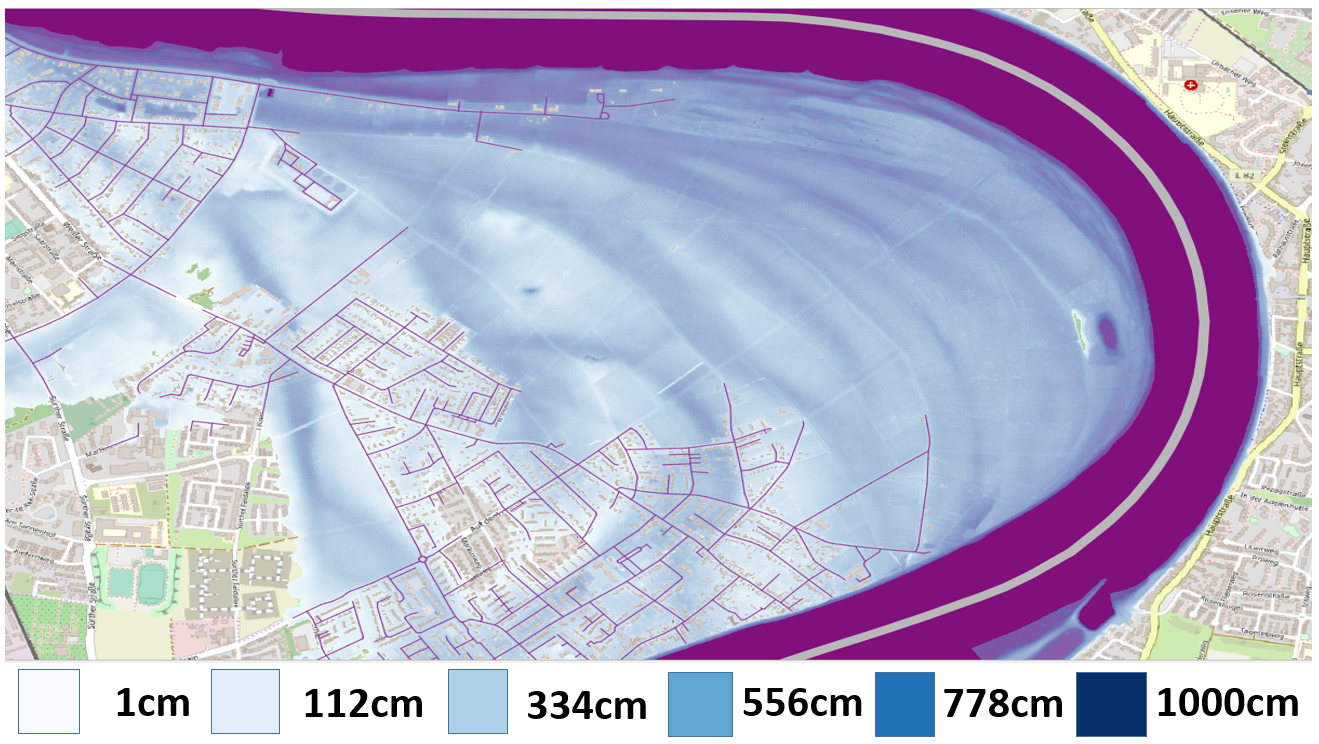}	
\caption{Floodmap depicting the flood altitude and a road network. The map legend informally describes the semantics of colors in terms of a fractional scale of flood altitudes.}
\label{fig:floodmap}
\end{subfigure}%
\hspace{0.2cm}
\begin{subfigure}{.48\textwidth}
	\centering
	\includegraphics[width=\linewidth]{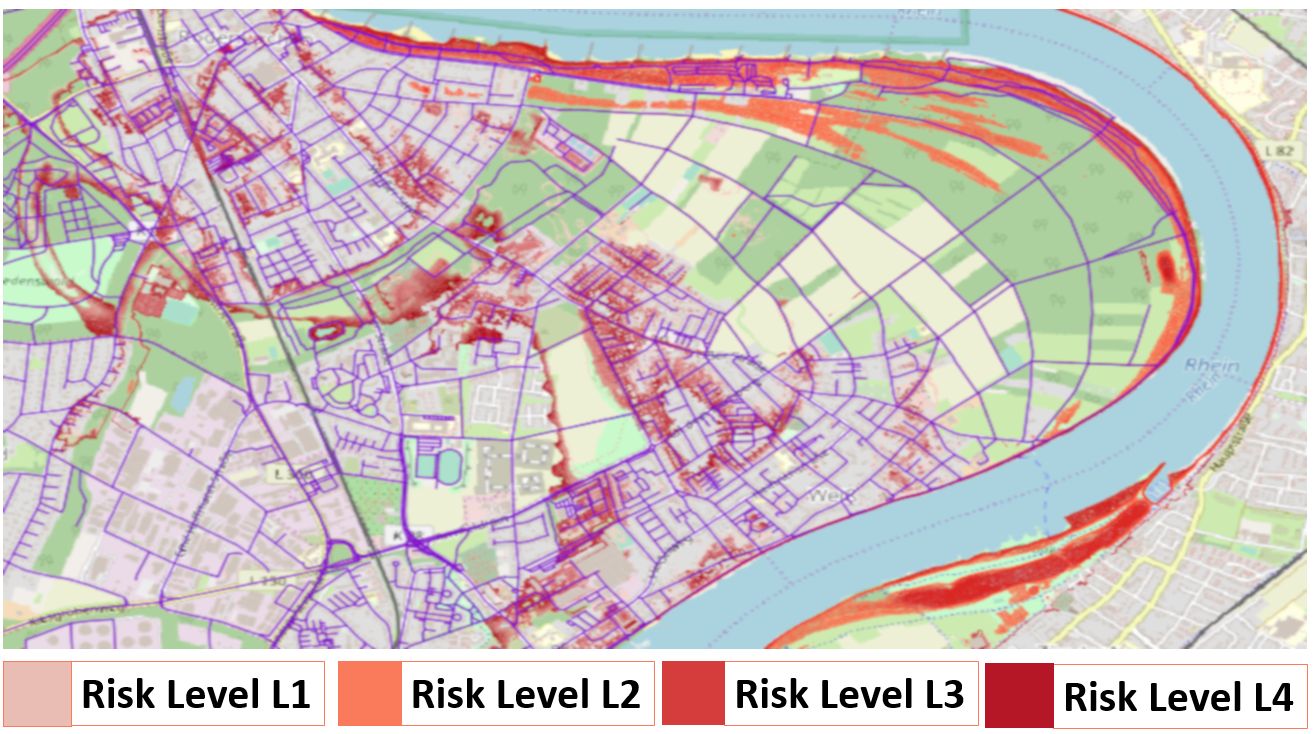}
\caption{Fire hazard risks displayed in different shades of red with darker shades implying higher risk levels. The map legend informally describes the risks using an ordinal scale.}
\label{fig:firemap}
\end{subfigure}
\caption{Visualizations of two sources of risk in Cologne}
\label{fig:hazardmaps}
\end{figure}

The Geospatial Semantic Web \cite{fonseca2008geospatial,van2019best} has grown in size and importance in the last decade. It is estimated that about 80\% of all data has a geospatial relation \cite{huxhold1991introduction}. Therefore, GeoSPARQL \cite{battle2012enabling} has been developed and became an OGC\footnote{\href{https://www.opengeospatial.org}{https://www.opengeospatial.org}} and W3C\footnote{\href{https://www.w3.org}{https://www.w3.org}} recommendation allowing for the representation and querying of geospatial data in the semantic web. GeoSPARQL and comparable approaches \cite{kyzirakos2012strabon,jena2019free} only provide support for geospatial vector data. However, geospatial data may also take the shape of a raster. It may, e.g., be obtained from aerial imagery or from simulation data to support tasks such as city planning and risk assessment as shown by the examples depicted in \Cref{fig:hazardmaps}.

Raster data must not be represented as vector geometries, because vector representations of raster data 
\begin{enumerate}
	\item are inefficient implying overconsumption of data storage. Raster data can be large and may be compressed efficiently.	
	\item are ineffective representations as they lack operations needed to query raster data e.g. raster algebra operations that transform raster data in ways not applicable to vector data.
	\item lack the semantics needed to appropriately represent raster data. Raster data is often visualized with RGB values, such as varying shades of blue for different flood altitudes. A semantic representation, however, should not represent color shades, but rather the underlying semantics, which should refer to data from the actual nominal, ordinal, interval or ratio scales and what they stand for.
\end{enumerate}
We propose GeoSPARQL+, an extension of the GeoSPARQL query language, and the  GeoSPARQL+ ontology in order to integrate geospatial raster data into the Semantic Web.

Let us consider the analysis of a flood as our running example. 
Our running example is depicted in \Cref{fig:floodmap} showing the overlay of two related datasets: 
\begin{enumerate}
	\item Vector data representing the roads of Cologne
	\item Raster data representing the altitudes of a simulated flood
\end{enumerate}
A query in one of our real-world use cases asks for all the road sections not covered by more than 10cm of water. This is only possible if the data model can represent raster data, vector data, semantics (road, water, depth, 10cm) and allows for joint querying of these representations.
Existing geographical information systems lack the explicit representation of semantics and require the user to manually adapt his high-level information need into a query of the low-level representation.
The GeoSPARQL standard \cite{battle2012enabling} and systems that currently support geographic information in the Semantic Web \cite{battle2012enabling,kyzirakos2012strabon,jena2019free,rdf4j2020,cerans2012graphical,graphdb2020,erling2012virtuoso} do not represent raster data, thus, they do not allow for asking such questions.
\definecolor{lightblue}{rgb}{247,251,255}
\definecolor{middleblue}{rgb}{95,166,209}
\definecolor{darkblue}{rgb}{8,48,107}
	\vspace{-0.5cm}
\noindent In the remainder of this paper, we will assume that there are data sources that contain vector data (e.g. roads in \Cref{fig:hazardmaps}) and raster data (e.g. flood altitudes \Cref{fig:floodmap} or fire hazards \cref{fig:firemap}). We describe a GeoSPARQL+ ontology which allows a data engineer to integrate these data into an RDF graph. A user may issue a semantic query against the RDF graph using GeoSPARQL+. To allow for these capabilities, this paper makes the following contributions:
\begin{enumerate}
	\item \emph{Semantic Representation of Raster Data:} A data model that allows for representing raster data and its semantics (\Cref{sec:raster2rdf}).
	\item \emph{GeoSPARQL Vocabulary Extension:} This vocabulary extension defines how to relate the raster data to semantic descriptions (\Cref{sec:vocabularyextension}).
	\item \emph{GeoSPARQL Query Language Extension:} A SPARQL extension which allows the interoperable use of semantic graph data, semantic vector geometries, and semantic raster data and uses map algebra \cite{tomlin1994map} to combine and modify rasters (\Cref{sec:geosparql2syntax,sec:geosparql2semdef}).
	\item \emph{Prototypical Implementation:} An open source implementation of the proposed approach for geospatial vector and raster data (\Cref{sec:implementation}).
	\item \emph{Requirements and Feasibility Check:} Deriving requirements of GeoSPARQL+ by discussing relevant use cases (\Cref{sec:usecasereqs}), assessing their feasibility and conducting a performance check of the implemented system (\Cref{sec:evaluation}).
\end{enumerate}
The tasks of data integration and visualization of query results are beyond the focus of this paper. More technical details about the supported functions may be found in our companion technical report \cite{homburg2020iswc}.
		\section{Foundations For Extending GeoSPARQL}
\label{sec:foundations}
In this publication we limit ourselves to 2D representations in order to remain concise. We see no major issue in extending our approach to higher dimensional representations.
We assume that all geographical representations relate to coordinate reference systems (CRS), as postulated in \cite{van2019best}. For conciseness of illustration we discard these relations and transformations between CRSs.
\subsection{Geometry}
\label{sec:geometry}
We formally define several OGC Simple Feature geometries \cite{herring2011opengis}, which we use in the remainder of this paper.
\begin{definition}{(Geometry)}
	A geometry $g \in \Geo$, with $\Geo$ representing the set of all geometries, is an instantiation of one of the following data structures:
	\begin{enumerate}
		\item A geometry $g$ may be a \emph{Point} $p=(x,y), p \in \mathbb{R}^2$, or
		\item A \emph{LineString} defined as a list of at least two different points denoted as $g=(p_0, p_1, \ldots, p_n), p_i \in \mathbb{R}^2$, or
		\item A \emph{Polygon} $g$ represented as a LineString with $g=(p_0, p_1, \ldots, p_n)$,  $p_0=p_n, p_i \in \mathbb{R}^2$ and all other points being unique. We further restrict ourselves to valid Polygons. In valid Polygons lines do not cross each other. A Polygon includes the encompassed area. 
		\item A geometry $g$ may also be a \emph{Rectangle}, which is a special polygon comprised of four LineStrings with the angles between connected LineStrings being $90^\circ$. $\Rect \subset \Geo$ is the set of all rectangles.
		\item Finally, a geometry may be a \emph{GeometryCollection} $g$, which itself is a finite set of geometries $g=\{g_1,\ldots,g_k\}, g_i \in \Geo$.
	\end{enumerate}
\emph{MultiPolygons} and \emph{MultiLineStrings} are examples of GeometryCollections.
\label{def:geometry}
\end{definition}
\noindent We assume that the function $\geometrytopointset:\Geo \rightarrow 2^{\mathbb{R}^2}$ exists which converts a geometry to a PointSet representation.
\subsection{RDF, SPARQL and GeoSPARQL}
\label{sec:definitions}
In order to semantically describe and query raster data we build upon the following standard definitions of SPARQL 1.1 \cite{world2013sparql,perez2006semantics}. 
%\todo{Mein Vorschlag: Hier eine Änderung des Wordings, dass wir nur die GeoSPARQL Funktionen formal definieren welche in der GeoSPARQL Definition zusätzlich zu SPARQL hinzukommen. Ich sehe keine Änderungen an den Standarddefinitionen die wir hier nicht markieren würden}
We provide the first formal definitions of the operators and filter functions which GeoSPARQL \cite{battle2012enabling} adds to the SPARQL query language and describe the resulting modified definitions of SPARQL 1.1 in the following.
In order to keep the definitions concise enough for this paper, we formalize syntax and semantics with 3 exemplary operators and 2 exemplary filter functions. We pick GeoSPARQL specific elements such that they are representative for the most common types of signatures.
The differences between SPARQL 1.1 and the GeoSPARQL extensions are {\color{blue} marked in blue fonts.}
\begin{definition}{(RDF Triple and RDF Graph)}
Let $I$, $B$ and $L$ be disjoint sets of IRIs, blank nodes and literals, respectively. An element of the set $(I \cup B) \times I \times (I \cup B \cup L)$ is called a \emph{triple} $t \in T$ with $T$ denoting the set of all triples. $G \in 2^{(I \cup B) \times I \times (I \cup B \cup L)}$ is called an \emph{RDF graph}. {\color{blue} $GL \subset L$ is the set of all geometry literals.} 
\end{definition}
\noindent In an RDF triple $(s, p, o)$, $s$, $p$ and $o$ are called \emph{subject}, \emph{predicate} and \emph{object}, respectively.
\noindent Geometry literals ($\GL$) are serialized according to the GeoSPARQL standard either as Well-Known-Text (WKT)\cite{wirz2004ogc} literals or as Geography Markup Language (GML)\cite{portele2007opengis} literals. 

\begin{definition}{(Triple Pattern)}
Let $V$ be a set of variables that is disjoint to $I$, $B$ and $L$. An element of $(I \cup B \cup L \cup V) \times (I \cup V) \times (I \cup B \cup L \cup V)$ is called a \emph{triple pattern}.
\end{definition}

\noindent The set of variables occurring in a triple pattern $\tp$ is abbreviated as $\var(\tp)$.

\begin{definition}{(Expression)}\label{def:expression}
An \emph{expression} is
\begin{tabbing}$Expression$ ::= \= \texttt{?X} \hspace*{12em} \= with $\texttt{?X} \in V$\\
 \> | $c$	\> with constant $c \in L \cup I$.\\
 \> | {\color{blue} $E_1 \cap E_2$} \> {\color{blue} with $E_1, E_2$ being expressions.}\\
 \> | {\color{blue} $\buffer(E_1, E_2, E_3)$}	\> {\color{blue} with $E_1, E_2, E_3$ being expressions}\\ 
 \> | {\color{blue} $\distance(E_1, E_2)$}	\> {\color{blue} with $E_1, E_2$ being expressions}
\end{tabbing}
\end{definition}

\begin{definition}{(Filter Condition)}\label{def:filterCondition}
A \emph{filter condition} is
\begin{tabbing}
$FilterCondition$ ::= \= \texttt{?X = c} \hspace*{10em} \= with $\texttt{?X} \in V$ and $c \in I \cup L$\\
 \> | \texttt{?X = ?Y}	\> with $\texttt{?X}, \texttt{?Y} \in V$\\
 \> | $\neg F$	\> with filter condition $F$\\
 \> | $F_1 \vee F_2$	\> with filter conditions $F_1$ and $F_2$\\
 \> | $F_1 \wedge F_2$	\> with filter conditions $F_1$ and $F_2$\\
 \> | {\color{blue} $E_1 \circled{$=$} E_2$}	\> {\color{blue} with $E_1, E_2$ being expressions} \\
 \> | {\color{blue} $E_1 \circled{$\cap$} E_2$}	\> {\color{blue} with $E_1, E_2$ being expressions}\\
\end{tabbing}
\end{definition}
\noindent {\color{blue}$\cap$, \circled{$=$}, \circled{$\cap$}} correspond to the GeoSPARQL operators  $\geointersection$,  $\geofequals$ and $\geointersects$ respectively \cite{open2012ogc}.
\noindent We provide complete list of all GeoSPARQL functions in our technical report that extends this paper \cite{homburg2020iswc}.
\begin{definition}{(Basic Graph Pattern)}
A \emph{basic graph pattern} ($BGP$) is
\begin{tabbing}
$BGP$ ::= \= $\tp$ \hspace*{80pt} \= a triple pattern $\tp$\\
 \> | $\{B\}$	\> a block of a basic graph pattern $B$\\
 \> | $B_1.B_2$	\> a conjunction of two basic graph patterns $B_1$ and $B_2$\\
 \> | $B~\texttt{FILTER}~F$	\> a filter pattern with $BGP$ $B$ and filter condition $F$\\
 \> | $B~\texttt{BIND}~( E~\texttt{AS}~\texttt{?X})$	\> a bind with $BGP$ $B$, expression $E$ and variable $\texttt{?X}$.
\end{tabbing}
\end{definition}

\begin{definition}{(Select Query)}
A \emph{select query} is defined as $\texttt{SELECT}~W~\texttt{WHERE}~B$ with $W \subseteq V$ and basic graph pattern $B$.
\end{definition}

%\noindent\textbf{SPARQL Semantics:}
% SPARQL semantics

\begin{definition}{(Variable Binding)}
A \emph{variable binding} is a partial function $\mu : V \pfun I \cup B \cup L$. The set of all variable bindings is $\Phi$.
\end{definition}

\noindent The abbreviated notation $\mu(\tp)$ means that variables in triple pattern $\tp$ are substituted according to $\mu$.

\begin{definition}{(Compatible Variable Binding)}
Two variable bindings $\mu_1$ and $\mu_2$ are \emph{compatible}, denoted by $\mu_1 \sim \mu_2$, if
\[\forall \texttt{?X} \in \dom(\mu_1) \cup \dom(\mu_2): \mu_1(\texttt{?X}) = \mu_2(\texttt{?X})\]
Thereby $\dom(\mu)$ refers to the set of variables of variable binding $\mu$.
\end{definition}

\begin{definition}{(Join)}
The \emph{join} of two sets of variable bindings $\Phi_1$, $\Phi_2$ is defined as \[\Phi_1 \bowtie \Phi_2 = \{ \mu_1 \cup \mu_2 | \mu_1 \in \Phi_1 \wedge \mu_2 \in \Phi_2 \wedge \mu_1 \sim \mu_2\}\]
\end{definition}

\begin{definition}{(Expression evaluation)}\label{def:expressionEvaluation}
The \emph{evaluation} of an expression $E$ over a variable binding $\mu$, denoted by $\eval{E}{\mu}$, is defined recursively as follows:
\begin{tabbing}
$\eval{\texttt{?X}}{\mu}$ \= := \= $\mu(\texttt{?X})$ \hspace*{90pt} \= with $\texttt{?X} \in V$.\\
$\eval{c}{\mu}$ \> := \> $c$ \> with $c$ being a constant, literal or IRI.\\
{\color{blue}$\eval{E_1 \cap E_2}{\mu}$ := $\eval{E_1}{\mu} \cap \eval{E_2}{\mu}$} \> \> \> {\color{blue} retrieves a geometry $g\in \Geo$ that represents}\\\>\>\> {\color{blue}all Points in the intersection of $\eval{E_1}{\mu},\eval{E_2}{\mu} \in \Geo$ \cite{battle2012enabling}}\\
{\color{blue}$\eval{\buffer(E_1,E_2,E_3)}{\mu}$ := $g$} \> \> \> {\color{blue} retrieves a bounding box $g \in \Rect$ of radius }\\
\> \> \> {\color{blue}$\eval{E_2}{\mu} \in \mathbb{R}$ around $\eval{E_1}{\mu} \in \Geo$ using the unit}\\ \> \> \> {\color{blue}given in $\eval{E_3}{\mu} \in I$ \cite{battle2012enabling}}  \\
{\color{blue}$\eval{\distance(E_1,E_2)}{\mu}:= c$} \> \> \> {\color{blue} returns the minimum distance $c \in \mathbb{R}$}\\ \> \> \> {\color{blue}between $\eval{E_1}{\mu} \in \Geo $ and $\eval{E_2}{\mu} \in \Geo$ \cite{battle2012enabling}} 
\end{tabbing}
\end{definition}

\begin{definition}{(Filter Condition Satisfaction)}\label{def:filterConditionSatisfaction}
Whether variable binding $\mu$ \emph{satisfies a filter condition} $F$, denoted by $\condeval{F}{\mu}$, is defined recursively as follows:
\begin{tabbing}
$\condeval{\texttt{?X} = c}{\mu}$ \hspace*{5em} \= holds if $\texttt{?X} \in \dom(\mu)$ and $\mu(\texttt{?X}) = c$.\\
$\condeval{\texttt{?X} = \texttt{?Y}}{\mu}$ \> holds if $\texttt{?X} \in \dom(\mu)$, $\texttt{?Y} \in \dom(\mu)$ and $\mu(\texttt{?X}) = \mu(\texttt{?Y})$.\\
$\condeval{\neg F}{\mu}$ \> holds if it is not the case that $\condeval{F}{\mu}$.\\
$\condeval{F_1 \vee F_2}{\mu}$ \> holds if $\condeval{F_1}{\mu}$ or $\condeval{F_2}{\mu}$.\\
$\condeval{F_1 \wedge F_2}{\mu}$ \> holds if $\condeval{F_1}{\mu}$ and $\condeval{F_2}{\mu}$\\
$\condeval{{\color{blue} E_1 \circled{$=$} E_2}}{\mu}$ \> {\color{blue}holds if $\eval{E_1}{\mu} \in \Geo$, $\eval{E_2}{\mu} \in \Geo$ and}\\  \> {\color{blue}$\geometrytopointset(\eval{E_1}{\mu}) = \geometrytopointset(\eval{E_2}{\mu})$}\\
$\condeval{{\color{blue} E_1 \circled{$\cap$} E_2}}{\mu}$ 	\> {\color{blue} holds if $\eval{E_1}{\mu} \in \Geo$, $\eval{E_2}{\mu} \in \Geo$ and}\\ \> {\color{blue}$\geometrytopointset(\eval{E_1}{\mu}) \cap \geometrytopointset(\eval{E_2}{\mu}) \neq \varnothing$}.
\end{tabbing}

\begin{comment}

$\condeval{\texttt{geo:equals}(E_1, E_2)}{\mu}$ \> holds if $\eval{E_1}{\mu} \in \mathbb{Geo}$, $\eval{E_2}{\mu} \in \mathbb{Geo}$ and\\  \> $\geometrytopointset(\eval{E_1}{\mu}) = \geometrytopointset(\eval{E_2}{\mu})$\\
$\condeval{\texttt{geo:intersects}(E_1, E_2)}{\mu}$ 	\> holds if $\eval{E_1}{\mu} \in \mathbb{Geo}$, $\eval{E_2}{\mu} \in \mathbb{Geo}$ and\\ \> $\geometrytopointset(\eval{E_1}{\mu}) \cap \geometrytopointset(\eval{E_2}{\mu}) \neq \varnothing$.
content...
\end{tabbing}
\end{comment}

\end{definition}

\begin{definition}{(SPARQL evaluation)}
The \emph{evaluation} of a SPARQL query $Q$ over an RDF graph $G$, denoted by $\eval{Q}{G}$, is defined recursively as follows:
\begin{tabbing}
$\eval{\tp}{G} := \{\mu | \dom(\mu) = \var(\tp) \wedge \mu(\tp) \in G\}$ \hspace*{10pt} \= with triple pattern $\tp$.\\
$\eval{\{B\}}{G} := \eval{B}{G}$ \> with basic graph pattern $B$.\\
$\eval{B_1.B_2}{G} := \eval{B_1}{G} \bowtie \eval{B_2}{G}$ \> with basic graph patterns $B_1$ and $B_2$.\\
$\eval{B~\texttt{FILTER}~F}{G} := \{\mu | \mu \in \eval{B}{G} \wedge \condeval{F}{\mu}\}$ \> with basic graph pattern $B$\\
				\> and filter condition $F$.\\
$\eval{B~\texttt{BIND}~( E~\texttt{AS}~\texttt{?X})}{G} := $ \> with basic graph pattern $B$,\\
\hspace*{10pt}$\{\mu \cup \{\texttt{?X} \mapsto \eval{E}{\mu}\} | \mu \in \eval{B}{G} \wedge \texttt{?X} \notin \dom(\mu) \}$	\> expression $E$ and variable $\texttt{?X}$.\\
$\eval{\texttt{SELECT}~W~\texttt{WHERE}~B}{G} := \{\mu_{|_W} | \mu \in \eval{B}{G}\}$ \> with basic graph pattern $B$ and $W \subseteq V$
\end{tabbing}
Thereby $\mu_{|_W}$ means that the domain of $\mu$ is restricted to the variables in $W$.
\end{definition}

\section{Use Case Requirements}
\label{sec:usecasereqs}
We now define requirements for use cases we have encountered when collaborating with companies developing geographical information systems.
\begin{enumerate}
	\item [U1] \textit{Client: Rescue Forces; Use case: Emergency rescue routing}\\ Rescue vehicles and routing algorithms guiding them need to know which roads are passable in case of flooding.\\
	Example query: \textit{"Give me all roads which are not flooded by more than 10cm"}	
	\item [U2] \textit{Client: Insurance; Use case: Risk assessment}\\ Insurances evaluate the hazard risk for streets and buildings in order to calculate the insurance premium.\\
	Example query: \textit{"Assess the combined risk of fire and flood hazards for all buildings in the knowledge base"}
	\item [U3] \textit{Client: Disaster Management Agency; Use case: Rescue capacity planning}\\ In case of disasters, the number of people present at a specified time and place needs to be estimated to prepare hospitals for casualties.\\
	Example query: \textit{"Give me the roads which contain elements at risk which are open to the public at 23rd May 2019 10.20am"}
	Note: An \emph{element at risk} is a term in disaster management describing a class of buildings affected by certain disasters \cite{homburg2020iswc}.
	\item [U4] \textit{Client: City Planning Agency; Use case: Rescue facility location planning}\\ Rescue forces should be stationed in a way that they can react fast to possible hazards and city planners should position rescue stations accordingly.\\
	Example query: \textit{"Give me the percentage of served hazardous areas, i.e. areas within a bounding box of 10km around a to-be-built rescue station at a given geocoordinate"}
\end{enumerate}
These example queries can currently not be expressed using GeoSPARQL. Abstracting from the given natural language query examples we have defined a graph data model for raster data and the syntax and semantics of GeoSPARQL+ that allow us to respond to these queries.
	    \section{Modeling Raster Data}
\label{sec:raster2rdf}
We have analyzed the requirements for representing raster data using examples like the ones depicted in \Cref{fig:hazardmaps} and use cases in \Cref{sec:usecasereqs}.
These examples show that we need to transform the following visual elements into semantic representations: 
\begin{enumerate}
	\item{Raster geometry}: A raster covers a geometrical area. In this paper, we limit ourselves to rectangular areas though other geometries might be supported in the future.
	\item{Atomic values}: In visualizations of raster data, atomic values are mapped onto pixel values. In simple cases this is a one-to-one mapping. Depending on the resolution of the raster and the rendered picture, it may also be a n:1 or 1:n or even an n:m mapping.
	\item{Atomic value geometry}:  Each atomic value represents the situation in a geometry area, typically in a rectangular area.
	\item{Raster legend}: A raster legend is a description of the semantic interpretation of the raster's atomic values. This description includes a categorical, ordinal, interval or fractional scale.
\end{enumerate} 
We formally define a raster following \cite{iso200519123} as:

\begin{minipage}{0.55\textwidth}
	\begin{definition}{(Raster)}
		Let R be the set of all rasters and $\mathbb{S}$ the set of all scales. A \emph{Raster} $r \in R$ is a partial function $r: \mathbb{R}^2 \pfun S$ which maps positions onto a scale $S \in \mathbb{S}$. We define a scale as a partially ordered set. In addition, every scale defines a NODATA value, a unique value which is not to be used elsewhere in the scale definition. The domain of a raster $dom(r)$ is the closed, rectangular region represented by its raster geometry for which its atomic values are defined. 
\end{definition}
dom(r) can be represented by a rectangle defined by its four corners ($p_l$,$p_b$,$p_r$,$p_t$),  
where $p_i=(x_i,y_i)$ and $x_l \leq x_r, x_l \leq x_t, x_l\leq x_b, x_r \geq x_t, x_r \geq x_b$ and
$y_l \geq y_b, y_t \geq y_r, y_l\leq y_t, y_b \leq y_r$.

Figure 2 shows an example of a raster.
\noindent In order to execute geometric operations on raster data and geometries we assume a function $\rastertogeom(r)$ returning $\dom(r)$ as a geometric object. \noindent In order to compare rasters to other rasters we assume an equality function. $\rastervaleq(r,r)$ compares the rasters atomic values and its domains.
\end{minipage} \hfill
\begin{minipage}{0.45\textwidth}
	%\begin{wrapfigure}{l}{4cm}

	\begin{turn}{10}
	\begin{tikzpicture}[scale=0.7]
	\tkzInit[xmax=6,ymax=6,xmin=0,ymin=0]
	%\tikz \draw (0,0) ellipse (2.5cm and 2.5cm);
	%\draw [fill=green,line width=7mm] (5.5,5.5) rectangle (0.5,0.5);
	\draw [white,dashed] (5.5,5.5) rectangle (0.5,0.5);
	%\node[left] {A} at (0,1) "ultra thick point";
	%\draw[red,dashed] (0.5,0.5) -- (5.5,0.5);
	%\draw[white,dashed] (5.5,0.5) -- (5.5,5.5);
	%\draw[yellow,dashed] (5.5,5.5) -- (0.5,5.5);
	%\draw[green,dashed] (0.5,0.5) -- (0.5,5.5);
	\node[draw,black,fill,circle,scale=0.2,text=black] at (6, 6)   (a) {$p_{t}$};
	\node[draw,white,scale=1,yshift=0.3cm,text=black] at (6, 6)   (a) {$p_{t}$};
	\node[draw,black,fill,circle,scale=0.2,text=black] at (6, 0)   (a) {$p_{r}$};
	\node[draw,white,scale=1,yshift=-0.3cm,text=black] at (6, 0)   (a) {$p_{r}$};
	\node[draw,black,fill,circle,scale=0.2,text=black] at (0, 0)   (a) {$p_{b}$};
	\node[draw,white,scale=1,yshift=-0.3cm,text=black] at (0, 0)   (a) {$p_{b}$};
	\node[draw,black,fill,circle,scale=0.2,text=black] at (0, 6)   (a) {$p_{l}$};
	\node[draw,white,scale=1,yshift=0.3cm,text=black] at (0, 6)   (a) {$p_{l}$};
	%\tikz\draw[very thick,blue] (0.5,0.5) -- node[point]{} (0.5,0.5) [point=red];
	\node[draw,white,scale=1.3,text=black,xshift=0.2cm,yshift=0.15cm] at (4, 4)   (a) {$c_{t}$};
	\node[draw,black,fill,circle,scale=0.2,text=black] at (4, 4)   (a) {$c_{t}$};
	\node[draw,white,scale=1.3,text=black,xshift=0.2cm,yshift=-0.2cm] at (4, 3)   (a) {$c_{r}$};
	\node[draw,black,fill,circle,scale=0.2,text=black] at (4, 3)   (a) {$c_{r}$};
	\node[draw,black,fill,circle,scale=0.2,text=black] at (4, 3)   (a) {$c_{b}$};
	\node[draw,black,fill,circle,scale=0.2,text=black] at (3, 3)   (a) {$c_{b}$};
	\node[draw,white,scale=1.3,text=black,xshift=-0.2cm,yshift=-0.2cm] at (3, 3)   (a) {$c_{b}$};
	\node[draw,black,fill,circle,scale=0.2,text=black] at (4, 4)   (a) {$c_{l}$};
	\node[draw,black,fill,circle,scale=0.2,text=black] at (3, 4)   (a) {$c_{l}$};
	\node[draw,white,scale=1.3,text=black,xshift=-0.2cm,yshift=0.6cm] at (3, 3)   (a) {$c_{l}$};
	\draw [step=1.0, very thick] (3,3) grid (4.0,4.0){};
	\node[draw,white,scale=1.3,text=black] at (3.5, 3.5)   (a) {$r_{i,j}$};
	\draw [step=2,densely dashed] (0,0) grid (6.0,6.0);
	\draw [densely dashed, step=1] (0,0) grid +(6.0,6.0);
	\draw [step=6] (0,0) grid (6.0,6.0);
	\end{tikzpicture}

	\end{turn}
		
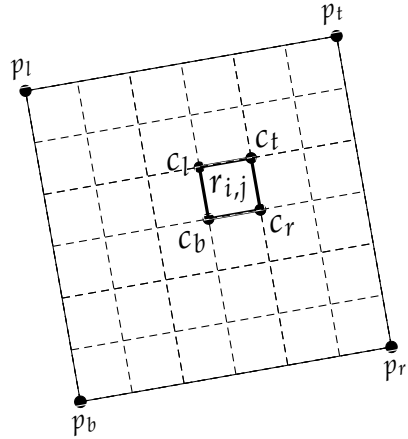
\captionof{figure}{Raster representation: The raster r is represented using a raster geometry $dom(r)$, a subdivision in cells and a scale $S \in \mathbb{S}$.}
	%\end{wrapfigure}
\end{minipage}
\begin{definition}{(Raster Literal)}
	The set $\RL \subset L$ with $\GL \cap \RL = \emptyset$ represents the set of all raster literals. 
\end{definition}
%\todo{Ich habe den NODATA value nun einfach als Teil der Scale definiert.}
\noindent We use the CoverageJSON format \cite{blower2017overview} to serialize rasters to raster literals, but many other textual serializations or even binary serializations are possible.
These representations assume that the raster geometry is divided uniformly into rectangular cell geometries (atomic value geometries in our previous definition).
A cell $c$ is a pair $(g,s) \in \Rect \times S$. We relate a cell $c$ to a raster $r$ via a pair of indexes $(i,j)$. $r_{i,j}$ refers to a specific cell indexed by $(i,j)$ in a raster $r$. $r_{i,j}(x,y)$ is undefined for values outside of the cell and has the identical value for all positions within the cell.
\noindent Thus, given $x,y$ such that $r_{i,j}(x,y)$ is defined, $c$ may be defined as 
$(raster2geom(r_{i,j}),r_{i,j}(x,y))$.

The function $\cellval: R \times \mathbb{R} \times \mathbb{R} \rightarrow \mathbb{R}$ retrieves the atomic value of a given raster cell. The function $\cellval2: R \rightarrow \{\mathbb{R}\}$ retrieves atomic values of all raster cells.

\textbf{Raster Algebra} or map algebra is a set based algebra to manipulate raster data.\label{sec:rasteralgebra}
Following \cite{tomlin1994map} we assume the definition of scale-dependent raster algebras with operations \circled{$\neg$}, $\oplus$ and $\circledless$ defined for the following signatures:

$$(1)\ \circled{$\neg$}: R \rightarrow R,\ (2)\ \oplus: R \times R \rightarrow R.\ (3)\ \circledless:  R \times \mathbb{R} \rightarrow R$$
\noindent The three operations we indicate here, their formal definitions given in \cite{tomlin1994map}, are examples for a  broader set of possible raster algebra operations. Most other algebraic operators exhibit the same signatures as one of these three example operations. Hence, syntax and semantics of other operators can be integrated into GeoSPARQL+ taking the integration of example operators as templates.

The \circled{$\neg$} function converts each atomic value different from 0 to 0, all 0 values to 1 and does not change NODATA values. The $\oplus$ function creates a new raster with the domain of the first raster. The resulting raster contains all values of the first raster which have no correspondence with the atomic values of the second raster (i.e. not map to the same position). All values with a correspondence are added together or ignored if one of the input values is the NODATA value of either of the two rasters. This function can be used to combine risks of fire and flood hazards given in two different rasters representing the same area. The $\circledless$ function takes one raster and one constant. It returns a new raster with the domain of the given raster. Atomic values smaller than the given constant are kept, all other values become the NODATA value. One application of this function is to only keep the flood altitude values displayed in \Cref{fig:floodmap} which signify an altitude value smaller than a given constant.

\noindent Implementations like PostGIS \cite{ramsey2005postgis} and JAI \cite{santos2004java} provide 26 and 108 raster functions respectively. Out of those we have implemented 14 in our system which we describe in \cite{homburg2020iswc}.
		\section{GeoSPARQL+}
\label{sec:geosparql20}
In order to describe raster data semantically,we must define (i) their geometries, (ii) their atomic values, (iii) the atomic value geometries, and (iv) the semantic meaning of raster's atomic values. The latter is specified in this section. When the raster's contents have been described, new functions are needed to filter, relate or modify the raster's atomic values in order to be useful in the application cases we would like to solve. Therefore we extend the GeoSPARQL query language to include such functions in \Cref{sec:geosparql2semdef,sec:geosparql2syntax}
\subsection{The GeoSPARQL+ Vocabulary}
\label{sec:vocabularyextension}
\begin{figure}[htbp]%[htbp]
	\centering
\includegraphics[width=\textwidth]{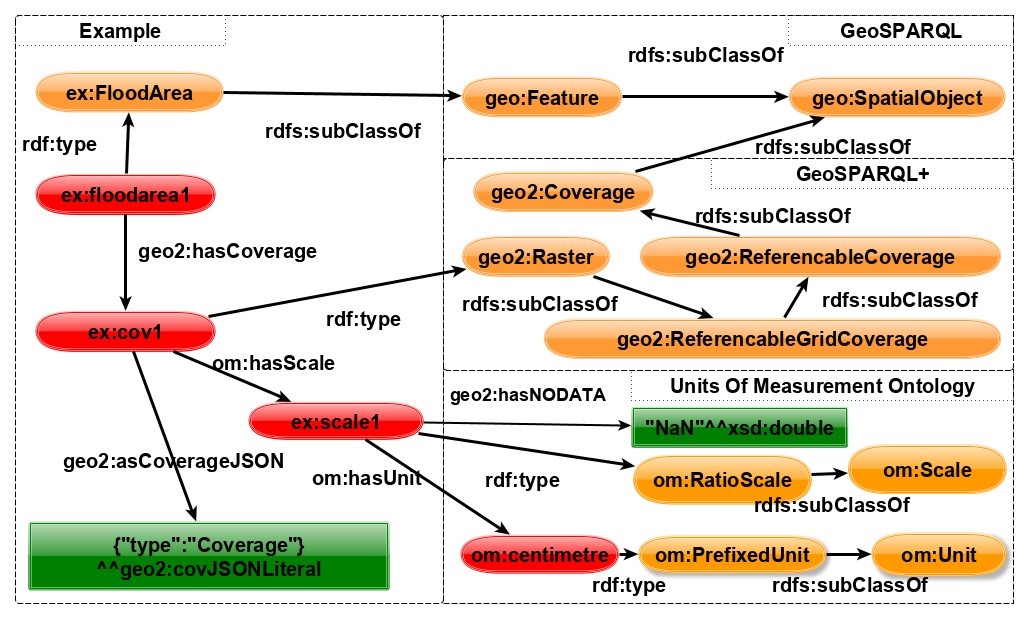}
	\caption{We use vocabularies of three different ontologies: The GeoSPARQL ontology describes the concepts \textit{geo:SpatialObject} and \textit{geo:Feature}, the OGC coverage hierarchy describes the abstract concepts of coverages and the unit of measurement vocabulary describes legends of raster data.}
	\label{fig:geosparql2raster}
\end{figure}

We define the new GeoSPARQL+ vocabulary (cf. \Cref{fig:geosparql2raster}). 
\vspace{-1cm}
\vspace{-0.7cm}
\\A raster is described by its semantic class (\emph{geo2:Raster}), and a scale which describes the semantic content of its atomic values.
In \Cref{fig:geosparql2raster}, we depict the example of a semantic class \textit{ex:FloodArea} which is assigned an instance of \textit{geo2:Raster} with a CoverageJSON literal (\Cref{lst:covjsonexample}) including the raster's type, a CRS, the raster's atomic values and their description. In order to re-use the representations of the CoverageJSON format, we model rasters in a concept hierarchy of OGC coverage types. By the OGC definition, a raster is a special type of coverage which is rectangular, i.e. a grid, and is georeferenced. This definition is reflected in \Cref{fig:geosparql2raster} in the given concept hierarchy. The instance of \textit{geo:Raster} connects to an instance of \textit{om:Scale} describing its legend and unit of measurement derived from the units of measurements ontology (UOM) as well as the scales NODATA value.

\hspace*{0.5cm}\begin{minipage}{0.96\linewidth}\begin{lstlisting}[label=lst:covjsonexample,caption=Coverage JSON Literal example]
{"type" : "Coverage","domain" : { "type" : "Domain", "domainType" : "Grid",
"axes": { "x" : { "values": [-10,-5,0] },"y" : { "values": [40,50] }
"referencing": [{"coordinates": ["y","x"],"system": {
"type": "GeographicCRS","id": "http://www.opengis.net/def/crs/EPSG/0/4979"}}]},
"observedProperty" : {
"ranges" : { "FloodAT" : { "type" : "NdArray", "dataType": "float", 
"axisNames": ["y","x"], "shape": [2, 2], "values" : [ 0.5, 0.6, 0.4, 0.6 ]}}}
\end{lstlisting}
\end{minipage}
\subsection{GeoSPARQL+ Syntax}
\label{sec:geosparql2syntax}
We added several new operators to the GeoSPARQL+ query language that allow to filter, modify and combine rasters as well as polygons. Due to space limitations, we present only one example for each of the three possibilites. A full list of the implemented functions is provided in \cite{homburg2020iswc}.
$\geometryintersection$ calculates intersections between arbitrary combinations of Geometries and Rasters, returning a Geometry. To get a raster as result instead, the $\coverageintersection$ can be used. $ \boxed{+}$ and $\boxed{<}$ provide two examples of raster algebra expressions.

GeoSPARQL+ defines the following new expressions to replace definition~\ref{def:expression}:
\begin{definition}{(GeoSPARQL+ Expression)}\label{def:geosparql2Expression}
		\begin{tabbing}
$Expression$ ::= \= \texttt{?X} \hspace*{12em} \= with $\texttt{?X} \in V$\\
| $c$	\> \> with constant $c \in L \cup I$. \\ 
%|$\hat{\cap}(E_1, E_2)$ \> \> with $E_1, E_2$ being expressions.\\
			 | {\color{blue}$\geometryintersection(E_1, E_2)$}\>	\>  {\color{blue} with $E_1, E_2$ being expressions}\\
			 | {\color{blue}$\rasterintersection(E_1, E_2)$} \>	\>  {\color{blue} with $E_1, E_2$ being expressions}\\
			 | {\color{blue}$E_1 \boxed{+} E_2$}\>	\> {\color{blue} with $E_1, E_2$ being expressions}\\
			 | {\color{blue}$E_1 \boxed{<} E_2$}\>	\>  {\color{blue} with $E_1, E_2$ being expressions}\\
			 | {\color{blue} $\boxed{\neg} E$ }\> \> {\color{blue} with $E$ being an expression}\\
			 | {\color{blue}$\rastertogeom(E)$}\>	\> {\color{blue} with $E$ being an expression}\\
			 | {\color{blue}$\rastervaleq(E_1, E_2)$}\>	\> {\color{blue} with $E_1, E_2$ being expressions}\\
			 | {\color{blue}$\geomtoraster(E_1, E_2)$}\>	\> {\color{blue} with $E_1, E_2$ being expressions}
		\end{tabbing}
\label{def:rasterfunctions}
\end{definition}
\noindent GeoSPARQL+ does not introduce new filter conditions in comparison to GeoSPARQL. However, the semantics of the previously defined filter conditions \circled{$=$} and \circled{$\cap$} are extended to also include raster literals.
\begin{comment}

 Raster-specific filter conditions extend the set of filter conditions defined for GeoSPARQL+.
GeoSPARQL+ defines the following new filter conditions in addition to definition~\ref{def:filterCondition}:
\begin{definition}{(GeoSPARQL+ Filter Condition)}\label{def:geosparql2FilterCondition}	
		\begin{tabbing}
	$FilterCondition$ ::= \= \texttt{?X = c} \hspace*{10em} \= with \texttt{?X}$ \in V$ and $c \in I \cup L$\\
	\> | \texttt{?X = ?Y}	\> with $\texttt{?X}, \texttt{?Y} \in V$\\
	\> | $\neg F$	\> with filter condition $F$\\
	\> | $F_1 \vee F_2$	\> with filter conditions $F_1$ and $F_2$\\
	\> | $F_1 \wedge F_2$	\> with filter conditions $F_1$ and $F_2$\\
	\> | $E_1 \circled{$=$} E_2$	\>  with $E_1, E_2$ being expressions\\
	\> | $E_1 \circled{$\cap$} E_2$	\>  with $E_1, E_2$ being expressions \\
	%$GeoFilterFunction$ ::= \= \hspace{10em} \= \\
	\>| {\color{blue} $\geoequals(E_1, E_2)$}\>  {\color{blue} with $E_1, E_2$ being expressions}\\
	%\>| {\color{blue} $\pointsetequals(E_1, E_2)$} \>  {\color{blue}with $E_1, E_2$ being expressions}\\
	\> | {\color{blue} $\intersects(E_1, E_2)$}\> {\color{blue} with $E_1, E_2$ being expressions}
\end{tabbing}
\label{def:geosparql20filtercondition}
\end{definition}
content...
\end{comment}
\subsection{GeoSPARQL+ Semantics}
\label{sec:geosparql2semdef}
We define the semantics of a GeoSPARQL+ expression in \Cref{def:geosparql2ExpressionEvaluation}.
In order to specify the intersection we map geometries and rasters to the corresponding PointSets. The result is a Geometry or Raster based on the selection of the user. In the special case of the intersection of two geometries, when a raster should be returned, we require a default value represented by parameter $E_3$ to which the atomic values of the created raster are mapped. The raster algebra functions $\coverageplus$ and $\coveragesmaller$ are mapped to their respective raster algebra expression defined in \Cref{sec:rasteralgebra}.

GeoSPARQL+ adds the following evaluations of expressions to definition~\ref{def:expressionEvaluation}:
\begin{definition}{(GeoSPARQL+ Expression Evaluation)}\label{def:geosparql2ExpressionEvaluation}	
	\begin{tabbing}
		%\hspace{10em} \hspace{9em} \= \hspace{12em} \= \\
		 $\eval{\geometryintersection(E_1, E_2)}{\mu}:=$ \= $\eval{E_1}{\mu} \cap \eval{E_2}{\mu}$\= \\\> if $\eval{E_1}{\mu}$ and $ \eval{E_2}{\mu} \in Geo$  \\
		$\eval{\geometryintersection(E_1, E_2)}{\mu}:=$\>$\eval{E_1}{\mu} \cap \rastertogeom(\eval{E_2}{\mu})$\\ \> if $\eval{E_1}{\mu} \in Geo$ and $\eval{E_2}{\mu} \in R$ \>  \\
		$\eval{\geometryintersection(E_1, E_2)}{\mu}:=\eval{\geometryintersection(E_2, E_1)}{\mu}$\\ \> if $\eval{E_1}{\mu} \in R$ and $\eval{E_2}{\mu} \in \Geo$ \>  \\
		$\eval{\rasterintersection(E_1, E_2)}{\mu} := r \in R$ with $\forall i,j: r_{i,j}=r1_{i,j}$ \\ \> if
		$\eval{E_1}{\mu}=r1,\eval{E_2}{\mu}=r2 \in R$\\  \> and $\dom(r1_{i,j})\cap \dom(r2_{i,j}) \neq \emptyset$\\
		$\eval{\coverageintersection(E_1, E_2)}{\mu}:= r \in R$ with $\forall i,j: r_{i,j}=r1_{i,j}$\\ \> if $\eval{E_1}{\mu}=r1 \in R$ and $\eval{E_2}{\mu}=g \in \Geo$ \\\> and  $\dom(r1_{i,j}) \cap g \neq \emptyset$ \\
		$\eval{\coverageintersection(E_1, E_2)}{\mu}:= \eval{\coverageintersection(E_2, E_1)}{\mu}$\\ \> if $\eval{E_1}{\mu}=r1 \in R$ and $\eval{E_2}{\mu}=g \in \Geo$ \>\\
		%$\eval{\rasterintersection(E_1, E_2,E_3)}{\mu} := r \in R$ with $\forall i,j: r_{i,j}=\dom(r1_{i,j})\cap \dom(r2_{i,j})\rightarrow \eval{E_3}{\mu}$ \\\> if $\eval{E_1}{\mu} \in \Geo, \eval{E_2}{\mu} \in \Geo, \eval{E_3}{\mu} \in \mathbb{R}$\\
		$\eval{\rastervaleq(E_1, E_2)}{\mu}:= r \in R$ \> with $\forall i,j: \dom(r1_{i,j})\cap \dom(r2_{i,j})\neq \emptyset$\\\> and $\cellval(r1_{i,j})==\cellval(r2_{i,j})$\\ \> if $\eval{E_1}{\mu}=r1, \eval{E_2}{\mu}=r2 \in R$ \>\\
		%$\eval{\coverageintersection(E_1,E_2,E_3)}{\mu}$ \> if $\eval{E_1}{\mu} \in Geo$ and $\eval{E_2}{\mu} \in Geo$\> \\
		% \>  with $r1_{i,j} \in \eval{E_1}{\mu}$ \\
		%\> $\forall r1_{i,j}: \dom(r1_{i,j})$\circled{$\cap$}$\eval{E_2}{\mu}$\\
		$\eval{\boxed{\neg} E}{\mu} := r \in R$ \> with $\forall i,j: r_{i,j}=\circled{$\neg$}r1_{i,j}$ if $\eval{E}{\mu}=r1 \in R$ \> \\	
		$\eval{E_1 \boxed{+} E_2}{\mu} := \eval{E_1}{\mu} \circled{+} \eval{E_2}{\mu}$ \> if $\eval{E_1}{\mu}$, $\eval{E_2}{\mu} \in R$ \\
		$\eval{E_1 \boxed{<} E_2}{\mu} := \eval{E_1}{\mu} \circled{<} \eval{E_2}{\mu}$ \> if $\eval{E_1}{\mu}, \eval{E_2}{\mu} \in R$ \\
		%$\eval{\rastertogeom(E)}{\mu} := \dom(\eval{E}{\mu})$ \> if $\eval{E}{\mu} \in R$\\
		 $\eval{\geomtoraster(E_1, E_2, E_3, E_4)}{\mu} :=r \in R$  with \\ \> $\forall (x,y) \in \buffer(\eval{E_1}{\mu},1,\uommeter) $\\\>$r(x,y)=\eval{E_2}{\mu}$\\\> if $\eval{E_1}{\mu} \in Geo, \eval{E_2}{\mu},\eval{E_3}{\mu},\eval{E_4}{\mu} \in \mathbb{R}$\\
		 \> with $\eval{E_3}{\mu}\cdot \eval{E_4}{\mu}$ indicating the number of cells
		  
	\end{tabbing}
\end{definition}
\noindent We define the semantics of a GeoSPARQL+ filter condition in \Cref{def:geosparql2FilterConditionEvaluation}.
The geo2:equals method returns true if two Raster or two Geometries are identical. The $\intersects$ method returns true if the PointSets of two Raster or Geometries overlap.
GeoSPARQL+ replaces the evaluation of the filter condition from definition~\ref{def:filterConditionSatisfaction} as follows:
\begin{definition}{(GeoSPARQL+ Filter Condition Satisfaction)}\label{def:geosparql2FilterConditionEvaluation}	
	\begin{tabbing}
		%\vspace{-5cm}
		%\hspace{10em} \= \hspace{10em} \= \\
		$\condeval{E_1 \circled{$=$} E_2}{\mu}$ \hspace{2em} \= holds if $\eval{E_1}{\mu},\eval{E_2}{\mu} \in \Geo$ and\\\> $\geometrytopointset(\eval{E_1}{\mu}) = \geometrytopointset(\eval{E_2}{\mu})$.\\
		$\condeval{E_1 \circled{$=$} E_2}{\mu}$ \> holds if $\eval{E_1}{\mu} \in R$ and $\eval{E_2}{\mu} \in \Geo$\\ \> and $\geometrytopointset(\rastertogeom(\eval{E_1}{\mu}))=\geometrytopointset(\eval{E_2}{\mu})$\\
		$\condeval{E_1 \circled{$=$} E_2}{\mu}$ \> holds if $\eval{E_1}{\mu} \in \Geo$ and $\eval{E_2}{\mu} \in R$ and $\condeval{E_2 \circled{$=$} E_1}{\mu}$\\
		$\condeval{E_1 \circled{$=$} E_2}{\mu}$ \> holds if $\eval{E_1}{\mu},\eval{E_2}{\mu} \in R$\\ \> and $\geometrytopointset(\rastertogeom(\eval{E_1}{\mu}))$\\ \>$=\geometrytopointset(\rastertogeom(\eval{E_2}{\mu}))$\\
		%$\eval{E_1}{\mu} \in Geo \wedge \eval{E_2}{\mu} \in Geo \wedge E_1=(p_0,...,p_n)=E_2$ \\
		$\condeval{E_1 \circled{$\cap$} E_2}{\mu}$ \> holds if $\eval{E_1}{\mu} \in R$, $\eval{E_2}{\mu} \in R$\\
		\> and $\geometrytopointset(\rastertogeom(\eval{E_1}{\mu}))$\\ \>$ \cap \geometrytopointset(\rastertogeom(\eval{E_2}{\mu})) \neq \varnothing$\\
		$\condeval{E_1 \circled{$\cap$} E_2}{\mu}$ \> holds if $\eval{E_1}{\mu} \in \Geo$, $\eval{E_2}{\mu} \in R$\\
		\> and $\geometrytopointset(\eval{E_1}{\mu}) \cap \geometrytopointset(\rastertogeom(\eval{E_2}{\mu})) \neq \varnothing$\\
		$\condeval{E_1 \circled{$\cap$} E_2}{\mu}$ \> holds if $\eval{E_1}{\mu} \in R$, $\eval{E_2}{\mu} \in \Geo$ and $\condeval{E_2 \circled{$\cap$} E_1}{\mu}$
	\end{tabbing}
\end{definition}
\paragraph{Further Functions} 
We have provided a couple of example functions and their signatures in order to show the principles of working with raster data. In practice, one needs a much larger set of functions and signatures. In particular the signatures \textit{geo:area}: $Geo \rightarrow \mathbb{R},$ \textit{geo2:max}: $R \rightarrow \mathbb{R} $ are used. \textit{geo:area} is a GeoSPARQL function calculating the area of a Geometry, \textit{geo2:max} calculates the maximum atomic value of a raster. 
We also use the additional raster algebra functions \textit{geo2:isGreater}: $R x \mathbb{R} \rightarrow \mathbb{R} $ and \textit{geo2:rasterUnion} $R x R \rightarrow \mathbb{R} $. The first one returns a raster which only includes atomic values greater than a given constant and the second one is the complement of the \textit{geo2:rasterIntersection} function.
		\section{Implementation}
\label{sec:implementation}
The implementation\footnote{\href{https://github.com/i3mainz/jena-geo}{https://github.com/i3mainz/jena-geo}} is built on Apache Jena \cite{jena2019free} and geosparql-jena \cite{albistongeosparql} and extends the ARQ query processor of Apache Jena with the GeoSPARQL+ functions defined in \cref{sec:geosparql20}. ARQ registers functions in an internal function registry which maps URIs to function implementations. The implementations were done in Java and used the Java Topology Suite library to implement vector geometry related functions, Apache SIS\footnote{\href{http://sis.apache.org}{http://sis.apache.org}} to represent rasters in Java and the Java Advanced Imaging Library (JAI) \cite{jaiswal2015spatial} to implement raster algebra operations. In addition, new literal types needed to be implemented in ARQ. geosparql-jena already provides support for vector literals (WKT and GML). To represent rasters we implemented CoverageJSON and Well-Known-Binary (WKB) literals with appropriate parsers for (de)serialization. In addition we implemented further functions defined in the SQL/MM standard \cite{stolze2003sql}. These functions help to prepare/modify vector geometries before they are compared or combined with rasters. Finally, we combined our implementation to work with a Apache Jena Fuseki triple store used for the feasibility study in \Cref{sec:evaluation}.

		\section{Feasibility}
\label{sec:evaluation}
\vspace{-0.1cm}
We work with the following datasets:
 	\vspace{-0.2cm}
\begin{enumerate}
	\item A vector dataset (GeoJSON): Road network of Cologne from OpenStreetMap
	\item A vector dataset (GeoJSON) of elements at risk extracted from OpenStreetMap
	\item Two rasters (flood altitude and fire hazards) of Cologne provided by a company simulating hazards
\end{enumerate}
The RDF graph contains the classes \textit{ex:Road}, classes for elements at risk and the classes \textit{ex:FloodRiskArea}, \textit{ex:FireRiskArea} for the rasters described in \Cref{sec:geosparql20}.
\subsection{GeoSPARQL+ Queries}
The feasibility check includes the four use cases defined in \Cref{sec:usecasereqs} and defines two queries per application case in GeoSPARQL+ and an equivalent query in SQL/MM \cite{stolze2003sql}. The GeoSPARQL+ query is executed on our prototypical implementation, the second query is executed on a POSTGIS implementation. For brevity we only illustrate the GeoSPARQL+ queries in \Cref{lst:floodaltitude,lst:riskassessment,lst:evacuationassessment,lst:cityplanning}.

\noindent The first query (\Cref{lst:floodaltitude}) solves usecase U1 and uses the raster algebra function \textit{geo:rasterSmaller} ($\boxed{<}$) (line 5) to filter those parts of a flood raster where roads that are still passable.\\
 \hspace*{0.4cm}\begin{minipage}{0.96\linewidth}
 	\begin{lstlisting}[label=lst:floodaltitude,caption={Use Case 1: Flood Altitude}]
 	SELECT ?road WHERE {
 	?road a ex:Road ; geo:hasGeometry ?roadseg . ?roadseg geo:asWKT ?roadseg_wkt .
 	?floodarea a ex:FloodRiskArea ; geo2:asCoverage ?floodarea_cov .
 	?floodarea_cov geo2:asCoverageJSON ?floodarea_covjson .
 	BIND(geo2:rasterSmaller(?floodarea_covjson,10) AS ?relfloodarea)
 	FILTER(geo2:intersects(?roadseg_wkt,?relfloodarea))}\end{lstlisting}
 \end{minipage}\\
 \noindent The second query (\Cref{lst:riskassessment}) solving use case U2 adds the values of two different rasters (fire and floodhazard) of the same area together (\textit{geo2:rasterPlus} ($\boxed{+}$) line 8) and extracts atomic values of the combined raster to assign a risk value to each given building. The maximum risk value per building is returned.
 
  \hspace*{0.4cm}\begin{minipage}{0.96\linewidth}
 	\begin{lstlisting}[label=lst:riskassessment,caption={Use case 2: Risk assessment}]
 	SELECT ?building (MAX(?riskvalue) AS ?riskmax) WHERE {
	?building a ex:Building ; geo:hasGeometry ?building_geom .
 	?building_geom geo:asWKT ?building_wkt .
 	?floodarea a ex:FloodRiskArea ; geo2:hasCoverage ?floodcov.
	?floodcov geo2:asCoverageJSON ?floodcov_covjson .
 	?firearea rdf:type ex:FireRiskArea ; geo2:hasCoverage ?firecov.
 	?firecov geo2:asCoverageJSON ?firecov_covjson .
 	BIND (geo2:rasterPlus(?firecov_covjson,?floodcov_covjson) AS ?riskarea) 
 	BIND (geo2:cellval2(geo2:rasterIntersection(?building_wkt,?riskarea)) AS ?riskvalue)
 	FILTER(geo2:intersects(?building_wkt,?riskarea))}\end{lstlisting}
 \end{minipage}\\
 \noindent The third query (\Cref{lst:evacuationassessment}) solving use case U3 combines the assessment of properties of vector geometries (line 10) with assessments gained from rasters (line 7) and GeoSPARQL functions like geo:buffer and geo:intersects (line 11-12) to evaluate roads with a higher priority to be evacuated. \\
 \hspace*{0.4cm}\begin{minipage}{0.96\linewidth}
 	\begin{lstlisting}[label=lst:evacuationassessment,caption={Use case 3: Rescue Capacity Planning}]
 	SELECT ?road WHERE{
	?road a ex:Road ; geo:hasGeometry ?roadgeom . ?roadgeom geo:asWKT ?road_wkt .
 	?ear a ear:ElementAtRisk ; geo:hasGeometry ?eargeom ; ex:openTime ?earopen ; ex:closeTime ?earclose .
 	?eargeom geo:asWKT ?ear_wkt .
 	?floodarea a ex:FloodRiskArea ; geo2:hasCoverage ?floodcov. ?floodcov geo2:asCoverageJSON ?floodcov_covjson .
 	?firearea rdf:type ex:FireRiskArea ; geo2:hasCoverage ?firecov. ?firecov geo2:asCoverageJSON ?firecov_covjson .
 	BIND (geo2:rasterPlus(?firecov_covjson,?floodcov_covjson) AS ?riskarea)
 	BIND("2019-05-23T10:20:13+05:30"^^xsd:dateTime AS ?givendate)
 	FILTER(?givendate>?earopen AND ?givendate<?earclose)
 	FILTER(geo:intersects(geo:buffer(?road_wkt,2,uom:meter),?ear))
 	FILTER(!geo:intersects(?road_wkt,?riskarea))}\end{lstlisting}
 \end{minipage}\\
Roads with a higher priority are near elements at risk for which we provide an ontology model in the appended technical report. The element at risk definition simplifies this query in comparison to an equivalent POSTGIS query, as the semantics are already explicitly stated.

\noindent Finally, the query for use case U4 (\Cref{lst:cityplanning}) combines the GeoSPARQL functions \textit{geo:area} (line 8) and \textit{geo:buffer} (line 7) with GeoSPARQL+ functions to intersect geometries and rasters (line 7-8) and to return a rasters geometry (line 8).

 \hspace*{0.5cm}\begin{minipage}{0.96\linewidth}
 	\begin{lstlisting}[label=lst:cityplanning,caption={Use case 4: City Planning}]
 	SELECT ?hazardcoveragepercentage WHERE {
 	?floodarea a ex:FloodRiskArea; geo2:hasCoverage ?floodcov.
 	?floodcov geo2:asCoverageJSON ?floodcov_covjson .
 	?firearea rdf:type ex:FireRiskArea ; geo2:hasCoverage ?firecov.
 	?firecov geo2:asCoverageJSON ?firecov_covjson .
 	BIND(geo2:rasterUnion(?firecov_covjson,?floodcov_covjson) AS ?hazardriskarea)
 	BIND(geo2:geometryIntersection(?hazardriskarea,geo:buffer(?locationtocheck,10,uom:km)) AS ?intersectarea)
 	BIND(geo:area(?intersectarea)/geo2:raster2geom(?hazardriskarea) AS ?hazardcoveragepercentage)
 	BIND("POINT(49.2,36.2)"^^geo:wktLiteral AS ?locationtocheck)}\end{lstlisting}
 \end{minipage}
 \subsection{Results}
 \label{sec:results}
 We measured the execution times of the introduced GeoSPARQL+ queries in comparison to equivalent SQL/MM \cite{wirz2004ogc} queries run on a POSTGIS implementation. The results are shown in \Cref{tab:queryresults}.
 \vspace{-0.3cm}
 \begin{table}
	\centering
\begin{tabular}{|c|c|c|}
	\hline
	Use case & GeoSPARQL+ & POSTGIS \\
	\hline
	Use case 1 & 112,423ms & 86,817ms\\
	\hline
	Use case 2 & 164,865ms & 108,357ms \\
	\hline
	Use case 3 & 134,865ms & 112,817ms \\
	\hline
	Use case 4 & 184,865ms & 140,357ms\\
	\hline
\end{tabular}
	\caption{Execution times of the given queries in the GeoSPARQL+ prototype vs. the comparison implementation in POSTGIS.}
	\label{tab:queryresults}
\end{table}
\vspace{-0.5cm}
\\ \Cref{tab:queryresults} shows that the execution time of our prototype is significantly longer than that of the native POSTGIS implementation. 
\subsection{Discussion}
\label{sec:discussion}
In \Cref{sec:geosparql20} have shown that the query solutions for use cases U1-U4 exploit different elements of GeoSPARQL+. Use case U1 relates a raster to a vector data set, use case U2 showcases the need of raster algebra operators to solve questions of combined risks, use case U3 combines values gained from rasters with attributes gained from vector data at the same geographic location. Both use case U2 and U3 make use of raster-aware filter functions. Finally, the query to solve use case U4 utilizes the raster to geometry function to create intersections between rasters with certain characteristics. 
We therefore illustrated the usefulness of GeoSPARQL+. 
Our prototypical implementation exhibits a slight performance decay between 23\% and 34\% for various example queries. We speculate that this degradation comes from overhead of dealing with semantics, lack of geospatial indices for rasters and further caches as well as a lack of technical optimizations that POSTGIS as a mature well-used system comes with. Considering that our implementation merely constitutes a proof of concepts, we consider this a graceful degradation and an acceptable result.
Future work may consider an improvement of its performance.
		\section{Related Work}
%\todo{Wieviel vom Related Work soll bleiben? Ich würde ja nur den Teil mit RDF-based stehen lassen}
\cite{koubarakis2010modeling} and \cite{perry2011sparql} proposed stSPARQL and SPARQL-ST, which extend SPARQL with spatiotemporal query capabilities for vector data. Spatiotemporal aspects for raster data and vector data are not considered by our approach but we see no major issues to combine the ideas of stSPARQL with our work. This is relevant as not only rasters with spatiotemporal aspects exist, but the content of raster data may also change over time.

Some approaches like LinkedGeoData \cite{auer2009linkedgeodata} convert SPARQL queries to SQL\\ queries in order to execute them on a native geospatial-aware SQL database. Similarly, hybrid systems such as Virtuoso \cite{erling2012virtuoso} add a semantic layer on top of a relational database such as POSTGIS \cite{ramsey2005postgis}. 
In principle, this would allow for accessing raster data, but has only been used to store and distribute vector data (cf. \cite{auer2009linkedgeodata}). We attribute this to a lack of semantic description of raster data which we address in this publication. Furthermore, we provide a solution independent of SQL datatabases and independent of the need for query conversions from SPARQL to SQL.

Relational spatial databases like POSTGIS \cite{ramsey2005postgis} or OGC geospatial webservices \cite{nogueras2005ogc} along with software suites such as QGIS\footnote{\href{https://qgis.org/de/site/}{https://qgis.org/de/site/}} and their accompanying libraries can handle, import, modify and query raster data, in particular with raster algebra. 
None of the aforementioned systems combines the advantages of linked data with the ability to semantically describe or access raster data information.

\noindent In addition to the previously mentioned work, there is a line of work that represents raster data as linked data (\cite{scharrenbach2012linked,quintero2009towards,world2014rdf}). These works do not consider how to query raster data. Hence, they lack the expressiveness required to cover our use cases. Similarly, \cite{bereta2018ontology} wrap raster data from a POSTGIS database and make it available as vector data that can be queried with GeoSPARQL. Because GeoSPARQL has no means for asking raster-specific queries (e.g. raster algebra), this work also lacks the expressiveness that our approach provides.

\noindent Another line of work includes representing and querying multi-dimensional arrays, SciSPARQL \cite{andrejev2015spatio}. While there is an overlap between managing raster data and arrays, raster data has geometric aspects that our approach supports (e.g. raster cell geometries, intersections and conversions between rasters and polygons, semantic descriptions of scales) that are not available when the underlying data model is restricted to arrays of real numbers. Hence, \cite{andrejev2015spatio} can not support our use cases, e.g. lacking intersecting street data and flooding data as we illustrate in \Cref{fig:floodmap}.
		\section{Conclusion}
\label{sec:conclusion}
We presented GeoSPARQL+ a novel approach that allows for the semantic description and querying of raster data in the semantic web. We expect these new capabilities to make publishing geospatial data in the geospatial semantic web more attractive and consider contributing this work to the currently discussed revision of GeoSPARQL \cite{homburg2020ogc,homburg2020ogc2}.
Future work could explore the semantic description of further OGC coverage types such as trajectories or even point clouds. Also, non-grid-based raster types should be investigated, as well as the representation of 3D rasters.

\noindent\textbf{Acknowledgements.} Work by Steffen Staab was partially supported by DFG through the project LA 2672/1, Language-integrated Semantic Querying (LISeQ).

		%\vspace{-0.3cm}
		\bibliography{references}
		\appendix
\section{GeoSPARQL function specifications}
In this section we list further GeoSPARQL function signatures to complete the formalization of GeoSPARQL for given expressions $E$. Functions returning a Bool value may be considered as filter functions. The function signatures are valid for the described inputs and are not defined for every other possible input. For example: The intersection of a geometry literal and a number is undefined.
\begin{longtable}{|c|c|}
	\hline
	Function Signature  & Function Description \\
	\hline
	\makecell{geof:boundary: $\Geo \rightarrow \Geo$} & Returns the boundary geometry of $E_1$ \\
	\hline
	\makecell{geof:convexHull: $\Geo \rightarrow \Geo$} & Conv($\eval{E_1}{\mu}$) \\
	\hline
	\makecell{geof:difference: $\Geo \rightarrow \Geo$} & Gets the difference area of two geometries \\
	\hline
	\makecell{geof:envelope: $\Geo \rightarrow \Rect$} & \makecell{Gets the minimum bounding\\ box of the given geometry} \\
	\hline
	\makecell{geof:getsrid: $\Geo \rightarrow \STR$} & Gets the srid number of $E_1$ \\
	\hline
	\makecell{geof:relate: $\Geo \rightarrow \mathbb{R}$} & \makecell{holds if $\eval{E_1}{\mu}$ and $\eval{E_2}{\mu}$ \\are spatially related\\ according to the DE-9IM matrix pattern} \\
	\hline
	\makecell{geof:symDifference: $\Geo \times \Geo \rightarrow \Geo$} & \makecell{Gets the symmetric difference\\ area of two geometries} \\
	\hline
	\makecell{geof:union: $\Geo \times \Geo \rightarrow \Geo$} & $\eval{E_1}{\mu} \cup \eval{E_2}{\mu}$ \\
	\hline
	\makecell{geof:sfCrosses: $\Geo \times \Geo \rightarrow BOOL$} & \makecell{holds if $\eval{E_1}{\mu} \cup \eval{E_2}{\mu}$  spatially crosses,\\ that is, the geometries have some,\\ but not all interior points in common.}  \\
	\hline
	\makecell{geof:sfContains: $\Geo \times \Geo \rightarrow BOOL$} & holds if $\eval{E_2}{\mu} \subseteq \eval{E_1}{\mu}$ \\
	\hline
	\makecell{geof:sfCovers: $\Geo \times \Geo \rightarrow BOOL$} & \makecell{holds if no point in Geometry $\eval{E_2}{\mu}$\\ is outside Geometry $\eval{E_1}{\mu}$} \\
	\hline
	\makecell{geof:sfDisjoint: $\Geo \times \Geo \rightarrow BOOL$} & holds if $\eval{E_1}{\mu} \cup \eval{E_2}{\mu}=\emptyset$ \\
	\hline
	\makecell{geof:sfOverlaps: $\Geo \times \Geo \rightarrow BOOL$} & \makecell{holds if $\eval{E_1}{\mu} \cup \eval{E_2}{\mu}\neq \emptyset $ \\and $\eval{E_1}{\mu} \cup \eval{E_2}{\mu}\neq \eval{E_1}{\mu}+\eval{E_2}{\mu}$} \\
	\hline
	\makecell{geof:sfTouches: $\Geo \times \Geo \rightarrow BOOL$} & \makecell{holds if the only points in\\ common between $\eval{E_1}{\mu} and \eval{E_2}{\mu}$\\ lie in the union of their boundaries.}  \\
	\hline
	\makecell{geof:sfWithin: $\Geo \times \Geo \rightarrow BOOL$} & holds if $\eval{E_1}{\mu} \supseteq \eval{E_2}{\mu}$ \\
	\hline
\end{longtable}
\section{SQL/MM function specifications}
In this section we list functions which are available in the SQL/MM standard \cite{stolze2003sql} to manipulate vector geometries. We also provide these functions in GeoSPARQL+, as they provide the ability to modify vector geometries in-query before they are related to rasters. We did not implement the full list of SQL/MM functions, but leave the implementations of the remaining functions to future work. However, we see no complications in implementing further SQL/MM functions.
\subsection{Vector Geometry Accessors}
\label{sec:vectorgeometryaccessors}
Vector geometry accessor functions access properties of or calculated attributes of a given geometry.
\begin{longtable}{|c|c|}
	\hline
	Function Signature  & Function Description \\
	\hline
	\makecell{$ST\_Centroid: \Geo \rightarrow \Geo$} & \makecell{Gets the centroid\\ of the given geometry} \\
	\hline
	\makecell{$ST\_EndPoint: \Geo \rightarrow \Geo$} & \makecell{Gets the last point\\ of the given geometry} \\
	\hline
	\makecell{$ST\_GeometryType: \Geo \rightarrow \STR$} & Gets the geometry type as String\\
	\hline
	\makecell{$ST\_HasRepeatedPoints: \Geo \rightarrow \BOOL$} &  holds if $E$ has repeated points \\
	\hline
	\makecell{$ST\_IsClosed: \Geo \rightarrow \BOOL$} & holds if $E$ is closed \\
	\hline
	\makecell{$ST\_IsCollection: \Geo \rightarrow \BOOL$} & holds if $E$ is a GeometryCollection \\
	\hline
	\makecell{$ST\_IsEmpty: \Geo \rightarrow \BOOL$} & holds if $E$ is an empty geometry \\
	\hline
	\makecell{$ST\_IsIsocelesTriangle: \Geo \rightarrow \BOOL$} & holds if $E$ is an isoceles triangle \\
	\hline
	\makecell{$ST\_IsMeasured: \Geo \rightarrow \BOOL$} & \makecell{holds if $E$ is a geometry\\ with an M coordinate} \\
	\hline
	\makecell{$ST\_IsPlanar: \Geo \rightarrow \BOOL$} & holds if $E$ is a 2D geometry \\
	\hline
	\makecell{$ST\_IsRectangle: \Geo \rightarrow \BOOL$} & holds if $E$ is a rectangle \\
	\hline
	\makecell{$ST\_IsSolid: \Geo \rightarrow \BOOL$} & holds if $E$ is a 3D geometry \\
	\hline
	\makecell{$ST\_IsSquare: \Geo \rightarrow \BOOL$} & holds if $E$ is a square \\
	\hline
	\makecell{$ST\_IsTriangle: \Geo \rightarrow \BOOL$} & holds if $E$ is a triangle \\
	\hline
	\makecell{$ST\_IsValid: \Geo \rightarrow \BOOL$} & holds if $E$ is a valid geometry \\
	\hline
	\makecell{$ST\_Length: \Geo \rightarrow \BOOL$} & Gets the length of the given geometry \\
	\hline
	\makecell{$ST\_M: \Geo \rightarrow \mathbb{R}$} & \makecell{Gets the first M coordinate\\ of the geometry} \\
	\hline
	\makecell{$ST\_MaxM: \Geo \rightarrow \mathbb{R}$} & \makecell{Gets the maximum M coordinate\\ of the geometry} \\
	\hline
	\makecell{$ST\_MaxX: \Geo \rightarrow \mathbb{R}$} & \makecell{Gets the maximum X coordinate\\ of the geometry} \\
	\hline
	\makecell{$ST\_MaxY: \Geo \rightarrow \mathbb{R}$} & \makecell{Gets the maximum Y coordinate\\ of the geometry} \\
	\hline
	\makecell{$ST\_MaxZ: \Geo \rightarrow \mathbb{R}$} & \makecell{Gets the maximum Z coordinate\\ of the geometry} \\
	\hline
	\makecell{$ST\_MinM(E_1): \Geo \rightarrow \mathbb{R}$} & \makecell{Gets the minimum M coordinate\\ of the geometry} \\
	\hline
	\makecell{$ST\_MinX(E_1): \Geo \rightarrow \mathbb{R}$} & \makecell{Gets the minimum X coordinate\\ of the geometry} \\
	\hline
	\makecell{$ST\_MinY(E_1): \Geo \rightarrow \mathbb{R}$} & \makecell{Gets the minimum Y coordinate\\ of the geometry} \\
	\hline
	\makecell{$ST\_MinZ(E_1): \Geo \rightarrow \mathbb{R}$} &\makecell{Gets the minimum Z coordinate\\ of the geometry} \\
	\hline
	\makecell{$ST\_NumDistinctPoints: \Geo \rightarrow \mathbb{N}$} & \makecell{Gets the number of distinct points\\ of the geometry} \\
	\hline
	\makecell{$ST\_NumPoints: \Geo \rightarrow \mathbb{N}$} & \makecell{Gets the number of points\\ of the geometry} \\
	\hline
	\makecell{$ST\_PointN: \Geo \times \mathbb{N} \rightarrow \Geo$} & \makecell{Gets the nth point\\ of the given geometry} \\
	\hline
	\makecell{$ST\_StartPoint: \Geo \rightarrow \Geo$} & \makecell{Gets the first point\\ of the given geometry} \\
	\hline
	\makecell{$ST\_X: \Geo \rightarrow \mathbb{R}$} & \makecell{Gets the first X coordinate\\ of the geometry} \\
	\hline
	\makecell{$ST\_Y: \Geo \rightarrow \mathbb{R}$} & \makecell{Gets the first Y coordinate\\ of the geometry} \\
	\hline
	\makecell{$ST\_Z: \Geo \rightarrow \mathbb{R}$} & \makecell{Gets the first Z coordinate\\ of the geometry} \\
	\hline
\end{longtable}
\subsection{Vector Geometry Transformations}
\label{sec:vectorgeometrytransformations}
Vector Geometry transformation functions transform a geometry to a different representation without changing the geometries coordinates.\\
\begin{longtable}{|c|c|}
	\hline
	Function Signature & Function Description \\
	\hline
	\makecell{$ST\_FlipCoordinates: \Geo \rightarrow \Geo$}  & \makecell{Flips the X and Y coordinates\\ of the given geometry} \\
	\hline
	\makecell{$ST\_Force2D: \Geo \rightarrow \Geo$} & \makecell{Converts the given geometry\\ to its 2D representation} \\
	\hline
	\makecell{$ST\_Force3D: \Geo \rightarrow \Geo$} &\makecell{ Converts the given geometry\\ to its 3D representation} \\
	\hline
	\makecell{$ST\_Force3DM: \Geo \rightarrow \Geo$} & \makecell{Converts the given geometry\\ to its 3DM representation} \\
	\hline
	\makecell{$ST\_Reverse: \Geo \rightarrow \Geo$} & \makecell{Reverses the coordinates\\ of the given geometry} \\
	\hline
	\makecell{$ST\_Scale: \Geo \times \mathbb{R} \rightarrow \Geo $} & Scales the geometry by a given factor \\
	\hline
	\makecell{$ST\_SimplifyPreserveTopology: \Geo \rightarrow \Geo$} & \makecell{Simplifies the geometry\\ while preserving its topology.} \\
	\hline
	\makecell{$ST\_Simplify: \Geo \rightarrow \Geo$} & \makecell{Simplifies the geometry according\\ to DouglasPeucker} \\
	\hline
\end{longtable}
\subsection{Vector Geometry Modifications}
\label{sec:vectorgeometrymodification}
Vector Geometry modification functions change a geometries coordinates by adding, removing or editing coordinates.
\begin{longtable}{|c|c|}
	\hline
	Function Signature & Function Description \\
	\hline
	\makecell{$ST\_AddGeometry: \Geo \times \Geo \rightarrow \Geo$} & \makecell{Adds a geometry to the\\ first given geometry} \\
	\hline
	\makecell{$ST\_AddPoint: \Geo \times \Geo \rightarrow \Geo$} & \makecell{Adds a point to the\\ first given geometry} \\
	\hline
	\makecell{$ST\_RemoveGeometry: \Geo \times \mathbb{N} \rightarrow \Geo$} & \makecell{Remove a geometry from the\\ first given geometry} \\
	\hline
	\makecell{$ST\_SetGeometry: \Geo \times \mathbb{N} x \Geo  \rightarrow \Geo$} & \makecell{Sets the nth geometry of the\\ first given geometry} \\
	\hline
	\makecell{$ST\_RemovePoint: \Geo \times \mathbb{N} \rightarrow Geo$} & \makecell{Removes the nth point\\ of a given geometry} \\
	\hline
	\makecell{$ST\_SetPoint: \Geo \times \mathbb{N} x \Geo \rightarrow \Geo$} & \makecell{Sets the nth point of\\ the given geometry} \\
	\hline
\end{longtable}
\subsection{Vector Geometry Exporters}
\label{sec:vectorgeometryexporters}
Vector Geometry export functions serialize geometries in vector literal formats defined in \cref{sec:vectorliteraltypes} and return a String representation of the literals value, i.e. an expression present in the set of all Strings $\STR$.
\begin{longtable}{|c|c|}
	\hline
	Function Signature & Function Description \\
	\hline
	\makecell{$ST\_AsGeoJSON: \Geo \rightarrow \STR$} & \makecell{Returns a GeoJSON representation\\ of the geometry} \\
	\hline
	\makecell{$ST\_AsGeoURI: \Geo \rightarrow \STR$} & \makecell{Returns a GeoURI representation\\ of the geometries' centroid}\footnote{\href{https://tools.ietf.org/html/rfc5870}{https://tools.ietf.org/html/rfc5870}} \\
	\hline
	\makecell{$ST\_AsGML: \Geo \rightarrow \STR$} & \makecell{Returns a GML representation\\ of the geometry} \\
	\hline
	\makecell{$ST\_AsKML: \Geo \rightarrow \STR$} & \makecell{Returns a KML representation\\ of the geometry} \\
	\hline
	\makecell{$ST\_AsOSMLink: \Geo \rightarrow \STR$} & \makecell{Returns a link to OpenStreetMap\\ pointing to a boundingbox of the given geometry} \\
	\hline
	\makecell{$ST\_AsTWKB: \Geo \rightarrow \STR$} & \makecell{Returns a TinyWKB representation\\ of the geometry}\footnote{\href{https://github.com/TWKB/Specification/blob/master/twkb.md}{https://github.com/TWKB/Specification/blob/master/twkb.md}} \\
	\hline
	\makecell{$ST\_AsWKT: \Geo \rightarrow \STR$} & \makecell{Returns a WKT representation\\ of the geometry} \\
	\hline
	\makecell{$ST\_AsWKB: \Geo \rightarrow \STR$} & \makecell{Returns a WKB representation\\ of the geometry} \\
	\hline
\end{longtable}
\subsection{Vector Geometry Relations}
\label{sec:vectorgeometryrelations}
Geometry relation functions relate two geometries in a 2dimensional space.
\begin{longtable}{|c|c|}
	\hline
	Function Signature & Function Description \\
	\hline
	\makecell{$ST\_CentroidDistance: \Geo \times \Geo \rightarrow \mathbb{R}$} & \makecell{Returns a GeoJSON representation\\ of the geometry} \\
	\hline
	\makecell{$ST\_FrechetDistance: \Geo \times \Geo \rightarrow \mathbb{R}$} & \makecell{Returns a GeoJSON representation\\ of the geometry}\cite{alt1995computing} \\
	\hline
	\makecell{$ST\_FullyWithinDistance: \Geo \times \Geo \times \mathbb{R} \rightarrow BOOL$} & \makecell{Holds if geometry $\eval{E_1}{\mu}$ is fully within\\ the distance given in $\eval{E_2}{\mu}$} \\
	\hline
	\makecell{$ST\_HausdorffDistance: \Geo \times \Geo \rightarrow Geo$} & \makecell{Returns the HausdorffDistance measure \\between two geometries}\cite{huttenlocher1993comparing} \\
	\hline
	\makecell{$ST\_WithinDistance: \Geo \times \Geo \times \mathbb{R} \rightarrow BOOL$} & \makecell{Holds if the two given geometries\\ are within the distance the third parameter.} \\
	\hline
\end{longtable}
\subsection{Additional Vector Literal Types}
\label{sec:vectorliteraltypes}
Additional vector literal types were added which may be used to export data in formats described in \Cref{sec:vectorgeometryexporters}
\begin{longtable}{|c|c|}
	\hline
	Literal Type & Example \\
	\hline
	GeoJSON Literal & "\{"geometry":"Point","coordinates":[0,0]\}" \\
	\hline
	GeoURI Literal & "geo:37.786971,-122.399677" \\
	\hline
	KML Literal & \makecell{"<kml xmlns="http://www.opengis.net/kml/2.2"><Placemark><Point>\\<coordinates>8.542952335953721,47.36685263064198,0</coordinates>\\</Point></Placemark></kml>"} \\
	\hline
	WKBLiteral Literal & \makecell{"010600000001000000010300000001000000050000007041F528CB332C413B509BE9\\710A594134371E05CC332C4111F40B87720A594147E56566CD332C419\\8DF5D7F720A594185EF3C8ACC332C41C03BEDE1710A59417041F528CB332C\\413B509BE9710A5941"}\\
	\hline
	TWKB Literal\footnote{\href{https://github.com/TWKB/Specification/blob/master/twkb.md}{https://github.com/TWKB/Specification/blob/master/twkb.md}} & "0x02000202020808"\\
	\hline
\end{longtable}
\section{GeoSPARQL+ function specifications}
This section lists further GeoSPARQL+ functions which have been proposed and implemented in our system. These functions operate on raster data.
\subsection{Raster Algebra Functions}
Raster algebra functions take at least one raster and either combines it with another raster, relates it to a given value or to itself.
\begin{longtable}{|c|c|}
	\hline
	Function Signature & Function Description \\
	\hline
	\makecell{$rasterAnd:R \times R \rightarrow R$} & $\forall r_{i,j} \in E_1, r2_{i,j} \in E_2 \rightarrow cellval(r_{i,j}) \& cellval(r2_{i,j})$ \\
	\hline
	\makecell{$rasterAndConst:R \times \mathbb{R} \rightarrow R$} & \makecell{$\forall r_{i,j} \in E_1 \rightarrow cellval(r_{i,j}) \& \eval{E_2}{\mu}$} \\
	\hline
	\makecell{$rasterDiv:R \times R \rightarrow R$} & \makecell{$\forall r_{i,j} \in E_1 and r2_{i,j} \in E_2 \rightarrow cellval(r_{i,j}) / cellval(r2_{i,j})$} \\
	\hline
	\makecell{$rasterDivConst:R \times \mathbb{R} \rightarrow R $} & \makecell{$\forall r_{i,j} \in E_1 \rightarrow cellval(r_{i,j}) / \eval{E_2}{\mu}$} \\
	\hline
	\makecell{$rasterEquals:R \times R \rightarrow R $}
	 & $\forall r_{i,j} \in E_1, r2_{i,j} \in E_2 \rightarrow cellval(r_{i,j}) == cellval(r2_{i,j})$\\
	\hline
	\makecell{$rasterEqualsConst:R \times \mathbb{R} \rightarrow R $}
	& \makecell{$\forall r_{i,j} \in E_1 \rightarrow cellval(r_{i,j}) == \eval{E_2}{\mu}$} \\
	\hline
	\makecell{$rasterExp(E_1,E_2) \rightarrow E_3$\\$ E_1,E_3 \in RL, E_2 \in \mathbb{R}$} &\makecell{$\forall r_{i,j} \in E_1 \rightarrow cellval(r_{i,j}) ^ \eval{E_2}{\mu}$}  \\
	\hline
	\makecell{$rasterGreater:R \times R \rightarrow R $} & \makecell{$\forall r_{i,j} \in E_1 \rightarrow cellval(r_{i,j}) > \eval{E_2}{\mu}$} \\
	\hline
	\makecell{$rasterInvert(E_1) \rightarrow E_2$\\$ E_1,E_2 \in RL$} & \makecell{$\forall r_{i,j} \in E_1 \rightarrow R_{i,j}*-1$} \\
	\hline
	\makecell{$rasterMax: R \rightarrow \mathbb{R}$} & \makecell{$\forall r_{i,j} \in E_1 \rightarrow max(R_{i,j})$} \\
	\hline
	\makecell{$rasterMean: R \rightarrow \mathbb{R}$} & \makecell{$\forall r_{i,j} \in E_1 \rightarrow mean(R_{i,j})$} \\
	\hline
	\makecell{$rasterMin: R \rightarrow \mathbb{R}$} &
	\makecell{$\forall r_{i,j} \in E_1 \rightarrow min(R_{i,j})$} \\
	\hline
	\makecell{$rasterMult: R \times R \rightarrow R$} & $\forall r_{i,j} \in E_1, r2_{i,j} \in E_2 \rightarrow cellval(r_{i,j}) * cellval(r2_{i,j})$ \\
	\hline
	\makecell{$rasterMultConst: R \times \mathbb{R} \rightarrow R$} & \makecell{$\forall r_{i,j} \in E_1 \rightarrow cellval(r_{i,j}) * \eval{E_2}{\mu}$} \\
	\hline
	\makecell{$rasterNot(E_1) \rightarrow E_2$\\$ E_1,E_2 \in RL$} & \makecell{$\forall r_{i,j} \in E_1 \rightarrow if\ cellval(r_{i,j})!=0\ then\ cellval(r_{i,j})=1$} \\
	\hline
	\makecell{$rasterOr: R \times R \rightarrow R$} & $\forall r_{i,j} \in E_1, r2_{i,j} \in E_2 \rightarrow cellval(r_{i,j}) || cellval(r2_{i,j})$ \\
	\hline
	\makecell{$rasterOrConst: R \times \mathbb{R} \rightarrow R$} &  \makecell{$\forall r_{i,j} \in E_1 \rightarrow cellval(r_{i,j}) || \eval{E_2}{\mu}$} \\
	\hline
	\makecell{$rasterPlus: R \times R \rightarrow R$} & $\forall r_{i,j} \in E_1, r2_{i,j} \in E_2 \rightarrow cellval(r_{i,j}) + cellval(r2_{i,j})$ \\
	\hline
	\makecell{$rasterPlusConst: R \times \mathbb{R}\rightarrow R$} & \makecell{$\forall r_{i,j} \in E_1 \rightarrow cellval(r_{i,j}) + \eval{E_2}{\mu}$} \\
	\hline
	\makecell{$rasterSmaller: R \times R \rightarrow R$} & \makecell{$\forall r_{i,j} \in E_1 \rightarrow cellval(r_{i,j}) < \eval{E_2}{\mu}$} \\
	\hline
	\makecell{$rasterSubtract: R \times R \rightarrow R$} & $\forall r_{i,j} \in E_1, r2_{i,j} \in E_2 \rightarrow cellval(r_{i,j}) - cellval(r2_{i,j})$ \\
	\hline
	\makecell{$rasterSubtractConst: R \times \mathbb{R} \rightarrow R$} & \makecell{$\forall r_{i,j} \in E_1 \rightarrow cellval(r_{i,j}) - \eval{E_2}{\mu}$} \\
	\hline
	\makecell{$rasterXor: R \times R \rightarrow R$} & \makecell{$\forall r_{i,j} \in E_1, r2_{i,j} \in E_2 \rightarrow cellval(r_{i,j}) xor cellval(r2_{i,j})$} \\
	\hline
	\makecell{$rasterXorConst: R \times \mathbb{R} \rightarrow R$} & \makecell{$\forall r_{i,j} \in E_1 \rightarrow cellval(r_{i,j})\ xor\ \eval{E_2}{\mu}$} \\
	\hline
\end{longtable}

\subsection{Raster Accessor Functions}
Raster Accessor functions provide access to specific raster content and raster attributes.
	\begin{longtable}{|c|c|}
		\hline
		Function Signature  & Function Description \\
		\hline
		\makecell{$rasterCell(E_1,E_2,E_3) \rightarrow E_4$\\$E_1 \in RL, E_2,E_3,E_4 \in \mathbb{R}$} & \makecell{Gets the atomic value of the cell\\ at the position given by $E_2,E_3$} \\
		\hline
		\makecell{$rasterCellHeight(E_1) \rightarrow E_2$\\ $E_1 \in RL, E_2 \in \mathbb{R}$}  & \makecell{Gets the height of a raster cell} \\
		\hline
		\makecell{$rasterCellWidth(E_1) \rightarrow E_2$\\$E_1 \in RL, E_2 \in \mathbb{R}$} & \makecell{Gets the width of a raster cell} \\
		\hline
		\makecell{$rasterEnvelope(E_1)\rightarrow E_2$\\$E_1 \in RL, E_2 \in \mathbb{R}$} & \makecell{Gest the raster geometry (Bounding Box)} \\
		\hline
		\makecell{$rasterHeight: R \rightarrow \mathbb{R}$} & \makecell{Gets the height of the raster} \\
		\hline
		\makecell{$rasterWidth: R \rightarrow \mathbb{R}$} & \makecell{Gets the width of the raster} \\
		\hline
\end{longtable}
\subsection{Raster Transformation Functions}
Raster transformation functions transform a given raster according to parameters given.
\begin{longtable}{|c|c|}
	\hline
	Function Signature & Function Description \\
	\hline
	\makecell{$rasterRescale(E_1,E_2,E_3) \rightarrow E_4$\\$E_1,E_4 \in RL, E_2,E_3 \in \mathbb{R}$} & \makecell{Resizes the raster to the width given in $E_2$\\ and the height given by $E_3$} \\
	\hline
	\makecell{$rasterResize(E_1,E_2,E_3) \rightarrow E_4$\\$E_1,E_4 \in RL, E_2,E_3 \in \mathbb{R}$} & \makecell{Resizes the raster to the width given in $E_2$\\ and the height given by $E_3$} \\
	\hline
\end{longtable}
\subsection{Raster To Vector Geometry Relation Functions}
Raster To Vector Geometry Relation Functions relate rasters to geometries by comparing the rasters' geometry to a vector geometry representation.
\begin{longtable}{|c|c|}
	\hline
	Function Signature & Function Description \\
	\hline
	\makecell{$ST\_rasterCoveredBy: R \times R \rightarrow BOOL$} & \makecell{Checks if a raster area\\ is covered by\\ other rasters geometry} \\
	\hline
	\makecell{$ST\_rasterCoveredBy: R \times Geo \rightarrow BOOL$} & \makecell{Checks if a raster area\\ is covered by\\ another geometry} \\
	\hline
	\makecell{$ST\_rasterCoveredBy: Geo \times R \rightarrow BOOL$} & \makecell{Checks if a raster area\\ is covered by\\ another geometry} \\
	\hline
	\makecell{$ST\_rasterEquals: R \times R \rightarrow BOOL$} & \makecell{$\eval{E_1}{\mu}==geom(\eval{E_2}{\mu}) if \eval{E_1}{\mu} \in R, \eval{E_2}{\mu} \in Geo$\\$geom(\eval{E_1}{\mu})==\eval{E_2}{\mu} if \eval{E_1}{\mu} \in Geo, \eval{E_2}{\mu} \in R$} \\
	\hline
	\makecell{$ST\_rasterEqualsContent: R \times R \rightarrow BOOL$} & \makecell{Checks if a raster area equals other\\ raster and the atomic values match}\\
	\hline
	\makecell{$ST\_rasterOverlaps: R \times R \rightarrow BOOL$} & \makecell{Checks if the raster areas overlap\\ with another rasters area} \\
	\hline
	\makecell{$ST\_rasterOverlaps: R \times Geo \rightarrow BOOL$} & \makecell{Checks if the raster areas overlap\\ with another geometry} \\
	\hline
	\makecell{$ST\_rasterOverlaps: Geo \times R \rightarrow BOOL$} & \makecell{Checks if the raster areas overlap\\ with another geometry} \\
	\hline
	\makecell{$ST\_rasterTouches: R \times R \rightarrow BOOL$} & \makecell{Checks if a rasters geometry touches\\ another rasters geometry} \\
	\hline
	\makecell{$ST\_rasterTouches: R \times Geo \rightarrow BOOL$} & \makecell{Checks if a rasters geometry touches\\ another geometry} \\
	\hline
	\makecell{$ST\_rasterTouches: Geo \times R \rightarrow BOOL$} & \makecell{Checks if a rasters geometry touches\\ another geometry} \\
	\hline
	\makecell{$ST\_rasterUnion: R \times R \rightarrow R$} & \makecell{Calculates the union of two\\ rasters} \\
	\hline
	\makecell{$ST\_rasterWithin: R \times R \rightarrow BOOL$} &\makecell{Checks if a raster area is within\\ other rasters area} \\
	\hline
	\makecell{$ST\_rasterWithin: R \times Geo \rightarrow BOOL$} &\makecell{Checks if a raster area is within\\ another geometry} \\
	\hline
	\makecell{$ST\_rasterWithin: Geo \times R \rightarrow BOOL$} &\makecell{Checks if a raster area is within\\ another geometry} \\
	\hline
	\makecell{$ST\_rasterWithinDistance: R \times R \rightarrow \BOOL$} &\makecell{ Checks if a raster area\\ is within distance\\ of the second given parameter} \\
	\hline
	\makecell{$ST\_rasterWithinDistance: R \times Geo \rightarrow \BOOL$} &\makecell{ Checks if a raster area\\ is within distance\\ of the second given parameter} \\
	\hline
	\makecell{$ST\_rasterWithinDistance: Geo \times R \rightarrow \BOOL$} &\makecell{ Checks if a raster area\\ is within distance\\ of the second given parameter} \\
	\hline
\end{longtable}
\subsection{Raster Exporter Functions}
\label{sec:rasterexporterfunctions}
Raster exporter functions export rasters to different serializations. The serializations are provided in raster literal types defined in \cref{sec:rasterlitearltypes}.
\begin{longtable}{|c|c|}
	\hline
	Function Signature & Function Description \\
	\hline
	\makecell{$ST\_asCoverageJSON: R \rightarrow \STR$}  &\makecell{Exports a raster to a CoverageJSON String} \\
	\hline
	\makecell{$ST\_asRasterWKB: R  \rightarrow \STR$} &\makecell{Exports a raster to RasterWKB}\footnote{\href{https://trac.osgeo.org/postgis/browser/trunk/raster/doc/RFC2-WellKnownBinaryFormat}{https://trac.osgeo.org/postgis/browser/trunk/raster/doc/RFC2-WellKnownBinaryFormat}} \\
	\hline
	\makecell{$ST\_asRasterHexWKB: R \rightarrow \STR$} &\makecell{Exports a raster to RasterWKB\footnote{\href{https://trac.osgeo.org/postgis/browser/trunk/raster/doc/RFC2-WellKnownBinaryFormat}{https://trac.osgeo.org/postgis/browser/trunk/raster/doc/RFC2-WellKnownBinaryFormat}} in hexadecimal form} \\
	\hline
\end{longtable}

\subsection{Raster Literal Types}
\label{sec:rasterlitearltypes}
We present further raster serialization types in this section. Rasters may be serialized as binary representations such as RasterWKB or as textual representations such as CoverageJSON or GMLCOV. Considering performance aspects, binary serializations might be preferrable over textual serializations.
\begin{longtable}{|c|c|}
	\hline
	Literal Type & Example \\
	\hline
	CoverageJSON Literal & "{"type" : "Coverage",...}" \\
	\hline
	RasterWKB Literal & \makecell{"00000000013FF00000000000003FF0000000000000000000000\\000000000000000000000000000000000000000000000\\0000000000000010E600020002040000010100"} \\
	\hline
\end{longtable}
\section{Elements At Risk Ontology Model}
\begin{figure}[htb]
	\includegraphics[width=\linewidth]{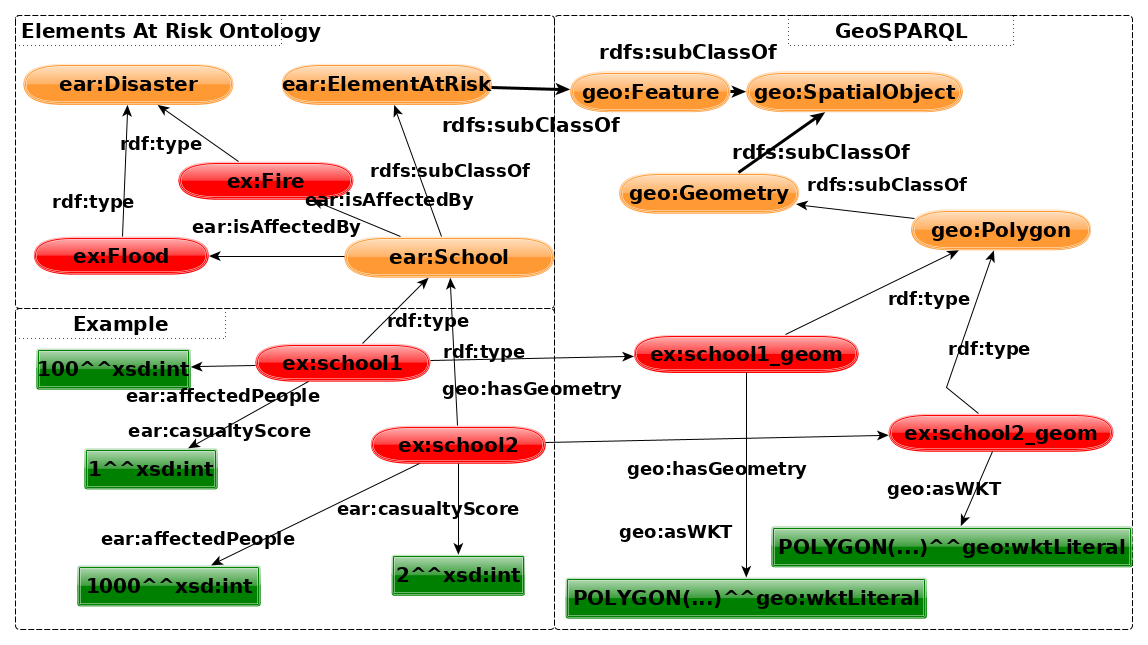}
	\caption{Elements At Risk Ontology}
	\label{fig:earontology}
\end{figure}

The ontology model shown in \cref{fig:earontology} represents elements at risk which are used in use case U3.
\\
An element at risk is a \textit{geo:Feature} connected to a \textit{geo:Geometry} susceptible to at least one type of disaster. Each instance of an element at risk is assigned a risk score. This risk score is calculated by combining a variety of parameters determining the severity of a risk situation. Examples of such parameters might be the amount of students of a school, the number of cars in an underground parking or the grade of protection of animals in a wildlife reserve. Depending on these parameters, determining the risk of a disaster and the capacities of mitigating the given disaster (e.g. emergency hospital beds vs. expected casualties at an element at risk) at a given place, the score will be assigned per instance. For example: If 100 students are present at a school to be rescued from a fire, 4 hospitals in the vicinity with an average capacity of 50 free beds need to be prepared to accommodate in the worst case 100 people with specific conditions like burns. If the school has a capacity of 1000 students, the capacity of the surrounding hospitals might not be sufficient, requiring the rescue forces to redistribute patients to clinics in other neighbouring cities. The first situation would be assigned a lower risk score, whereas the latter situation would be assigned a higher risk score.\\
In the context of GeoSPARQL+, the existence of such an ontology structure and the modeling of this knowledge allows a user to use GeoSPARQL+ to formulate a more concise query as compared to an SQL query. In SQL, the dependencies of risk types are usually distributed over a variety of tables, which firstly should be joined and secondly filtered before they are combined with raster representations. While the execution time may not be too different, semantics allow for a more easy formulation of the query.
\begin{comment}
\subsection{Raster Modification Functions}
Raster Modification functions modify contents of rasters without applying a raster algebra function.
\begin{longtable}{|c|c|c|c|}
	\hline
	Function Signature & Parameters & Return Type & Function Description \\
	\hline
	$ST\_rasterUnion(E_1,E_2)$ & $E_1,E_2\in GL \cup RL$ & GL & Calculates the union of a raster/geometry combination \\
	\hline
\end{longtable}
\subsection{Raster Relation Functions}
Raster relation functions describe relations between rasters which cannot be applied to geometries.
\begin{longtable}{|c|c|c|c|}
	\hline
	Function Signature & Parameters & Return Type & Function Description \\
	\hline
	$ST\_rasterUnion(E_1,E_2)$ & $E_1,E_2\in GL \cup RL$ & GL & Calculates the union of a raster/geometry combination \\
	\hline
\end{longtable}
content...
\end{comment}

\end{document}